\documentclass[pra,floats,aps,nofootinbib,superscriptaddress]{revtex4}
\usepackage{graphicx}
\usepackage{amssymb}
\usepackage{epsfig}
\input psfig.sty 
\begin{document}
\title{Approximate analysis of search algorithms with ``physical'' methods}
\author{S. Cocco}
\affiliation{CNRS-Laboratoire de Dynamique des Fluides Complexes,
3 rue de l'universit\'e 67084 Strasbourg, France}
\author{R. Monasson}
\affiliation{CNRS-Laboratoire de Physique Th{\'e}orique de l'ENS,
24 rue Lhomond, 75005 Paris, France}
\affiliation{CNRS-Laboratoire de Physique Th{\'e}orique,
3 rue de l'universit\'e, 67084 Strasbourg, France}
\author{A. Montanari}
\affiliation{CNRS-Laboratoire de Physique Th{\'e}orique de l'ENS,
24 rue Lhomond, 75005 Paris, France}
\author{G. Semerjian}
\affiliation{CNRS-Laboratoire de Physique Th{\'e}orique de l'ENS,
24 rue Lhomond, 75005 Paris, France}

\begin{abstract}
An overview of some methods of statistical physics applied to the analysis
of algorithms for optimization problems (satisfiability of Boolean
constraints, vertex cover of graphs, decoding, ...) 
with distributions of random inputs is proposed.
Two types of algorithms are analyzed: complete procedures with
backtracking (Davis-Putnam-Loveland-Logeman algorithm) and incomplete,
local search procedures (gradient descent, random walksat, ...).
The study of complete algorithms makes use of physical concepts such as
phase transitions, dynamical renormalization flow, growth processes, ...
As for local search procedures, the connection between computational
complexity and the structure of the cost function landscape is questioned,
with emphasis on the notion of metastability. 
\end{abstract}

\maketitle

\section{Introduction}

The computational effort needed to deal with large combinatorial
structures considerably varies with the task to be performed and the
resolution procedure used\cite{papadimi}.  The worst case complexity
of a task, more precisely an optimization or decision problem, is
defined as the time required by the best algorithm to treat any
possible inputs to the problem. For instance, the sorting problem of a
list of $N$ numbers has worst-case complexity $\sim N\log N$: there
exists several algorithms that can order any list in at most $\sim N\log
N$ elementary operations, and none with asymptotically
less operations.  Unfortunately, the worst-case complexities
of many important computational problems, called NP-Complete, is not
known. Partitioning a list of $N$ numbers in two sets with equal
partial sums is one among hundreds of such NP-complete problems.  It
is a fundamental conjecture of theoretical computer science that there
exists no algorithm capable of partitioning any list of length $N$,
or of solving any other NP-Complete problem with inputs of size $N$,  in
a time bounded by a polynomial of $N$. Therefore, when dealing with 
such a problem, one necessarily uses algorithms which may takes
exponential times on some inputs. Quantifying how `frequent' these
hard inputs are for a given algorithm is the question answered by the
analysis of algorithms. In this paper, we will present an overview
of recent works done by physicists to address this point, and more
precisely to characterize the average performances, called hereafter
complexity, of a given
algorithm over a distribution of inputs to an optimization problem.

The history of algorithm analysis by physical methods/ideas is at least as
old as the use of computers by physicists. One well-established chapter in
this history is, for instance, the analysis of Monte Carlo sampling algorithms
for statistical mechanics models. In this context, it is well known that
phase transitions, {\em i.e.} abrupt changes in the physical 
properties of the model, can imply a dramatic increase in the time
necessary to the sampling procedure. This phenomenon is commonly
known as critical slowing down. The physicists' insight in this problem
comes mainly from the analogy between the dynamics of algorithms and the 
physical dynamics of the system. This analogy is quite natural:
in fact many algorithms mimick the physical dynamics itself.

A quite new idea is instead to abstract from physically motivated problems
and use statistical mechanics ideas for analyzing the dynamics of algorithms.
In effect there are many reasons which suggest that analysis of algorithms and 
statistical physics should be considered close relatives. In
both cases one would like to understand the asymptotic behavior
of dynamical processes acting on exponentially large (in the size
of the problem) configuration spaces. The differences between the two
disciplines mainly lie in the methods (and, we are tempted to say, the style)
of investigation. Theoretical computer science derives rigorous results
based on probability theory. However these results are sometimes too weak 
for a complete characterization of the algorithm. Physicists provide instead
heuristic results based on intuitively sensible approximations. These
approximations are eventually validated by a comparison with numerical
experiments. In some lucky cases, approximations are asymptotically
irrelevant: estimates are turned into conjectures left for future rigorous
derivations. 

Perhaps more interesting than stylistic differences is the {\it point of
view} which physics brings with itself. Let us highlight two consequences of
this point of view.

First, a particular importance is attributed to ``complexity phase
transitions'' {\em i.e.}  abrupt changes in the resolution
complexity as some parameter defining the input distribution
is varied\cite{AI,Friedgut}. 
We shall consider two examples in the next Sections:
\begin{itemize}
\item Random Satisfiability of Boolean constraints (SAT).
In $K$-SAT one is given an instance, that is, a
 set of $M$ logical constraints (clauses) 
among $N$ boolean variables, and wants to find a truth assignment for the
  variables which fulfill all the constraints. Each clause is the logical OR of
$K$ literals, a literal being one of the $N$ variables or its
  negation e.g. $(x_1 \vee x_{17} \vee \overline{x_{31}})$ for 3-SAT. 
Random $K$-SAT is the $K$-SAT problem supplied with a distribution of 
inputs uniform over all instances having fixed values of $N$ and $M$. The
limit of interest is $N,M\to \infty$ at fixed ratio $\alpha=M/N$ of
clauses per variable\cite{Mit,Hans}. 
\item Vertex cover of random graphs (VC).
An input instance of the VC decision problem consists in a
graph $G$ and an integer number $X$.
The problem consists in finding a way to distribute $X$ covering marks over the 
vertices in such a way that every edge of the graph is
covered, that is, has at least one of its ending vertices marked.
A possible distribution of inputs is provided by drawing random graphs
$G$ {\em \`a la} Erd\"os-Reny\`i {\em i.e.}
with uniform probability among all the graphs having
$N$ vertices and $E$ edges. The limit of interest is
$N,E\to\infty$ at fixed ratio $c=2E/N$ of edges per vertex. 
\end{itemize}
The algorithms for random SAT and VC we shall consider in the next
Sections undergo a complexity phase transition as the input parameter
$\pi$ ($=\alpha$ for SAT, $c$ for VC) crosses some critical threshold
$\pi_{\rm alg}$. Typically resolution of a randomly drawn
instance requires linear time below the threshold
$\pi<\pi_{\rm alg}$ and exponential time above 
$\pi>\pi_{\rm alg}$. The observation that most difficult
instances are located near the phase boundary confirms the relevance of
the phase-transition phenomenon.

Secondly, a key role is played by the intrinsic (algorithm
independent) properties of the instance under study. The intuition is
that, underlying the dramatic slowing down of a particular algorithm,
there can be some {\it qualitative} change in some structural property
of the problem e.g. the geometry of the space of solutions.  While
there is no general understanding of this question, we can further
specify the above statements case-by-case. Let us consider, for
instance, a local search algorithm for a combinatorial optimization
problem. If the algorithm never increases the value of the cost
function $F(C)$ where $C$ is the configuration (assignment) of variables to be
optimized over, the number and geometry of the local minima of $F(C)$
will be crucial for the understanding of the dynamics of the algorithm. This
example is illustrated in Sec. \ref{gradxor}.  While the ``dynamical''
behavior of a particular algorithm is not necessarily related to any
``static'' property of the instance, this approach is nevertheless of
great interest because it could provide us with some `universal'
results. Some properties of the instance, for example, may imply the
ineffectiveness of an entire class of algorithms.

While we shall mainly study in this paper
the performances of search algorithms applied to
hard combinatorial problems as SAT, VC, we will also consider easy, that is,
polynomial problems as benchmarks for these algorithms.
The reason is that we want to understand if the average hardness of 
resolution of solving NP-complete problems with a given distribution
of instances and a given algorithm truly reflects the
intrinsic hardness of these combinatorial problems or is
simply due to some lack of
efficiency of the algorithm under study. The benchmark problem we shall
consider is random XORSAT.
It is a version of a satisfiability problem, much simpler than SAT
from a computational complexity point of
view\cite{Crei}. The only but essential
difference with SAT is that a
clause is said to be satisfied if the exclusive, and not inclusive, 
disjunction  of its literals is true. 
XORSAT may be recast as a linear algebra problem, where a set of
$M$ equations involving $N$ Boolean variables must be satisfied modulo
2, and is therefore solvable in polynomial time by
various methods e.g. Gaussian elimination. Nevertheless, it is
legitimate to ask what are the performances of general search algorithms
 for this kind of polynomial computational problem. In particular,
we shall see that some algorithms requiring exponential times to solve
random SAT instances behave badly on random XORSAT instances too.
A related question we shall focus on in Sec.~\ref{CodeSection} is
decoding, which may also, in some cases, be expressed as the
resolution of a set of Boolean equations.

The paper is organized as follows. In Sec.~\ref{DpllSection} we shall review
backtracking search algorithms, which, roughly speaking, work in the
space of instances. We explain the general ideas and then illustrate them on
random SAT (Sec.~\ref{DpllSatSection}) and VC (Sec.~\ref{DpllVcSection}).  
In Sec.~\ref{DpllFlucSection} we consider the fluctuations in running times of
these algorithms and analyze the possibility of exploiting these fluctuations
in random restart strategies. In Sec.~\ref{LocalSection} we turn to local
search algorithms, which work in the space of configurations.
We review the analysis of such algorithms for decoding problems 
(Sec.~\ref{CodeSection}), random XORSAT (Sec.~\ref{gradxor}),
and SAT (Sec.~\ref{WalkSatSection}).
Finally in the Conclusion we suggest some possible future developments 
in the field.

\section{Analysis of the Davis-Putnam-Loveland-Logeman search procedure}

\subsection{Overview of the algorithm and physical concepts}
\label{DpllSection}

\begin{figure}
\centerline{\includegraphics[scale=0.38,angle=-90]{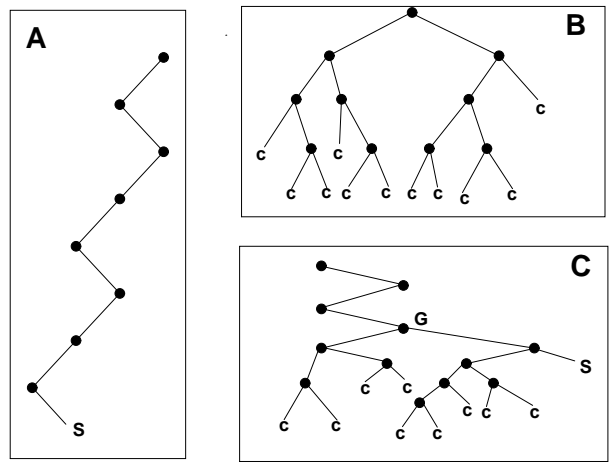}}
\caption{Types of search trees generated by the DPLL solving procedure
for variables taking $v=2$ values at most. 
Nodes (black dots) stand for the choices of variables made by the
heuristic, and edges between nodes denote the elimination of
unitary clauses. 
{\bf A.} {\em simple branch:} the algorithm finds
easily a solution without ever backtracking. {\bf B.} {\em dense tree:}
in the absence of solution, the algorithm builds a ``bushy'' tree,
with many branches of various lengths, before stopping.  {\bf C.} {\em
mixed case, branch + tree:} if many contradictions arise before
reaching a solution, the resulting search tree can be decomposed in a
single branch followed by a dense tree. The junction G is the highest
backtracking node reached back by DPLL.}
\label{trees}
\end{figure}

In this section, we briefly review the Davis-Putnam-Loveland-Logemann
(DPLL) procedure~\cite{DP,survey}. A decision problem can be formulated as a
constrained satisfaction problem, where a set of variables must be
sought for to fulfill some given constraints. For simplicity, we
suppose here that variables may take a finite set of values with
cardinality $v$ e.g. $v=2$ for SAT or VC. DPLL is an exhaustive search
procedure operating by trials and errors, the sequence of which can be
graphically represented by a search tree (Fig.~\ref{trees}). The tree
is defined as follows: {\bf (1)} A node in the tree corresponds to a
choice of a variable.  {\bf (2)} An outgoing branch (edge) codes for
the value of the variable and the logical implications of this choice
upon not yet assigned variables and clauses.  Obviously a node gives
birth to $v$ branches at most.  {\bf (3)} Implications can lead to:
{\bf (3.1)} a violated constraint, then the branch ends with $C$
(contradiction), the last choice is modified (backtracking of the
tree) and the procedure goes on along a new branch (point 2 above);
{\bf (3.2)} a solution when all constraints are satisfied, then the
search process is over; {\bf (3.3)} otherwise, some constraints remain
and further assumptions on the variables have to be done (loop back to
point 1).

A computer independent measure of computational complexity, that is,
the amount of operations necessary to solve the instance, is given by
the size $Q$ of the search tree {\em i.e.} the number of nodes it
contains. Performances can be improved by designing sophisticated
heuristic rules for choosing variables (point 1).  The resolution time
(or complexity) is a stochastic variable depending on the instance
under consideration and on the choices done by the variable assignment
procedure. Its average value, $\bar Q$, is a function of the input
distribution parameters $\pi$ e.g. the ratio $\alpha$ of clauses per
variable for SAT, or the average degree $c$ for the VC of random
graphs, which can be measured experimentally and that we want to calculate
theoretically.  More precisely, our aim is to determine the values of the
input parameters for which the complexity is linear, $\bar Q=\gamma
\,N$ or exponential, $\bar Q = 2^{N\,\omega}$, in the size $N$ of the
instance and to calculate the coefficients $\gamma, \omega$ as
functions of $\pi$.

The DPLL algorithm gives rise to a dynamical process.  Indeed, the
initial instance is modified during the search through the assignment
of some variables and the simplification of the constraints that
contain these variables. Therefore, the parameters of the input
distribution are modified as the algorithm runs. This dynamical
process has been rigorously studied and understood in the case of a
search tree reducing to one branch (tree A in
Figure~\ref{trees})\cite{fra2,Fra,Achl,Fri,Kir,kir2}.  Study of trees
with massive backtracking e.g. trees B and C in Fig.~\ref{trees} is
much more difficult.  Backtracking introduces strong correlations
between nodes visited by DPLL at very different times, but close in
the tree.  In addition, the process is non Markovian since instances
attached to each node are memorized to allow the search to resume
after a backtracking step.

\begin{figure}
\begin{center}
\includegraphics[height=250pt,angle=0]{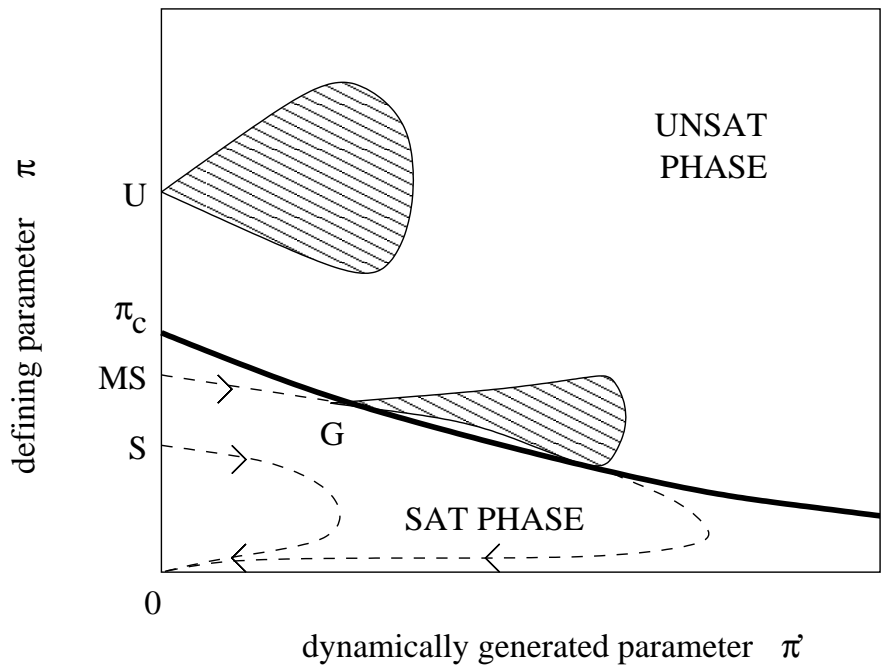}
\end{center}
\caption{Schematic representation of the resolution trajectories in the
sat (branch trajectories symbolized by dashed lines) and unsat (tree 
trajectories represented by hatched regions) phases. For simplicity
we have considered the case where both $\pi$ and $\pi'$ are scalar
and not vectorial parameters. Vertical
axis is the instance distribution defining parameter $\pi$. Instances
are almost always satisfiable if $\pi < \pi _c$, unsatisfiable if
$\pi > \pi _c$. Under the action of DPLL, the distribution of
instances is modified and requires another parameter $\pi'$ to be
characterized (horizontal axis), equal to, say, zero prior to any action
of DPLL. For non zero values of $\pi'$, the critical value
of the defining parameter $\pi$ obviously changes; the line
$\pi _c (\pi')$ defines a boundary separating typically sat from unsat
instances (bold line).
When the instance is unsat (point U), DPLL takes an exponential time
to go through the tree trajectory. For satisfiable and easy
instances, DPLL goes along a branch trajectory in a linear time
(point S). The mixed case
of hard sat instances (point MS) 
correspond to the branch trajectory crossing the boundary separating
the two phases (bold line), which leads to the exploration of unsat subtrees
before a solution is finally found.}
\label{schemoins}
\end{figure}

The study of the operation of DPLL is based on the following, elementary 
observation. Since instances are modified when treated by DPLL,
description of their statistical properties 
generally requires additional parameters 
with respects to the defining parameters $\pi$ of the input distribution. 
Our task therefore consists in
\begin{enumerate}
\item identifying these extra parameters $\pi'$\cite{kir2};
\item deriving the phase diagram of this new, extended distribution 
$\pi,\pi'$ to identify, in the $\pi,\pi'$ space, the critical surface 
separating instances having solution with high probability
(satisfiable phase) from instances having generally
no solution (unsatisfiable phase), see Fig.~\ref{schemoins}.
\item tracking the evolution of an instance under resolution with time
$t$ (number of steps of the algorithm), that is, the trajectory of its 
characteristic parameters $\pi(t),\pi'(t)$ in the phase diagram.
\end{enumerate}
Whether this trajectory remains confined to one of the two phases or
crosses the boundary inbetween has dramatic consequences on the 
resolution complexity. We find three average behaviours, schematized
on Fig.~\ref{schemoins}:
\begin{itemize}
\item if the initial instance has a solution and the trajectory
remains in the sat phase, resolution is typically linear,
and almost no backtracking is present (Fig.~\ref{trees}A).
The coordinates of the trajectory $\pi(t),\pi'(t)$ of the instance
in the course of the resolution obey a set of coupled ordinary
differential equations accounting for the changes in the distribution
parameters done by DPLL.
\item if the initial instance has no solution, 
solving the instance, that is, finding a proof of unsatisfiability,
takes exponentially large time and makes use of
massive backtracking (Fig.~\ref{trees}B). Analysis of the search tree
is much more complicated than in the linear regime, and requires a partial
differential equation that gives information on the population of
branches with parameters $\pi,\pi'$ throughout the growth of the search tree.
\item in some intermediary regime, instances have solutions but finding
one requires an exponentially large time (Fig.~\ref{trees}C).
This may be related to the crossing of the boundary between sat and unsat
phases of the instance trajectory. We have therefore a mixed
behaviour which can be understood through combination of the two above
cases.
\end{itemize}
We now explain how to apply concretly this approach to the cases of
random SAT and VC.

\subsection{Average analysis of the Random SAT problem}
\label{DpllSatSection}

\begin{figure}
\centerline{\includegraphics[scale=0.5,angle=0]{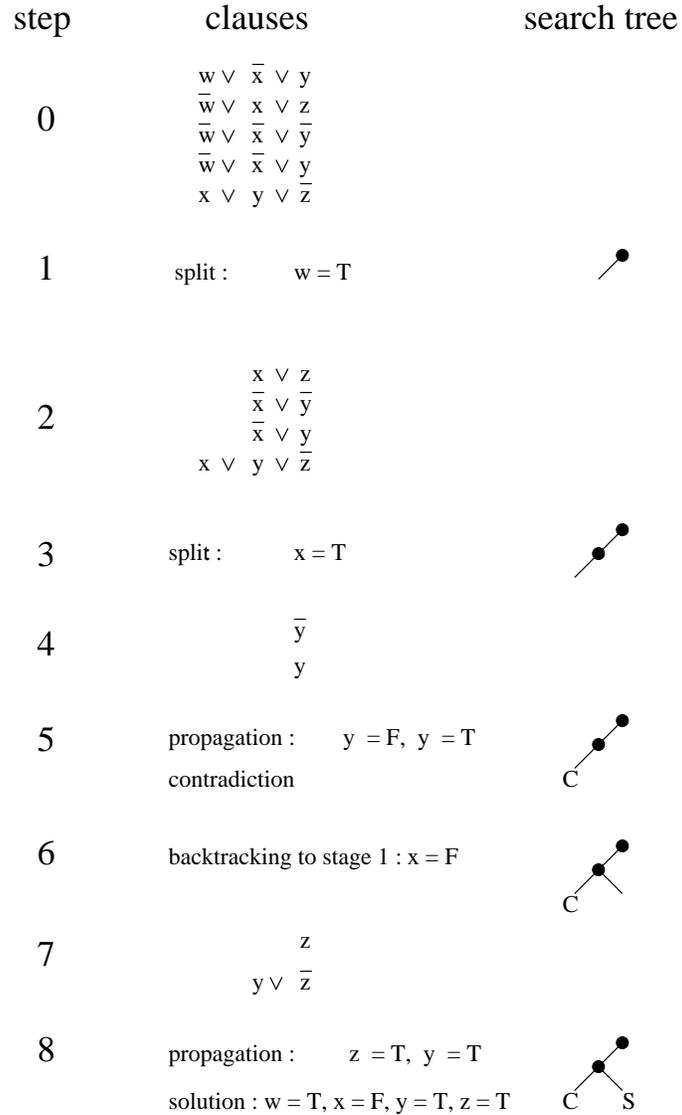}}
\caption{Example of 3--SAT instance and Davis-Putnam- Loveland-Logemann
resolution.
{\bf Step~0.}  The instance consists of $M=5$ clauses
involving $N=4$ variables $x,y,w,z$, which can be assigned to true (T) or false
(F). $\bar w$ means (NOT $w$) and $\vee$ denotes the logical OR. The search
tree is empty.  {\bf 1.}  DPLL randomly selects a clause among the
shortest ones, and assigns a variable in the clause 
to satisfy it, e.g. $w=$T 
(splitting with the Generalized Unit Clause --GUC-- heuristic \cite{fra2}).
A node and an edge symbolizing respectively the variable chosen ($w$)
and its value (T) are added to the tree.  {\bf 2.}  The logical
implications of the last choice are extracted: clauses containing $w$
are satisfied and eliminated, clauses including $\bar w$ are
simplified and the remaining ones are left unchanged. If no unitary
clause ({\em i.e.} with a single variable) is present, a new choice of
variable has to be made.  {\bf 3.}  Splitting takes over. Another node
and another edge are added to the tree.  {\bf 4.}  Same as step 2 but
now unitary clauses are present.  The variables they contain have to
be fixed accordingly.  {\bf 5.}  The propagation of the unitary
clauses results in a contradiction. The current branch dies out and
gets marked with C.  {\bf 6.}  DPLL backtracks to the last split
variable ($x$), inverts it (F) and creates a new edge.  {\bf 7.}  Same
as step 4.  {\bf 8.}  The propagation of the unitary clauses
eliminates all the clauses. A solution S is found and the instance is 
satisfiable.  For an
unsatisfiable instance, unsatisfiability is proven when backtracking
(see step 6) is not possible anymore since all split variables have
already been inverted. In this case, all the nodes in the final search
tree have two descendent edges and all branches terminate by a
contradiction C.}
\label{algo}
\end{figure}

\begin{figure}
\begin{center}
\includegraphics[height=400pt,angle=-90]{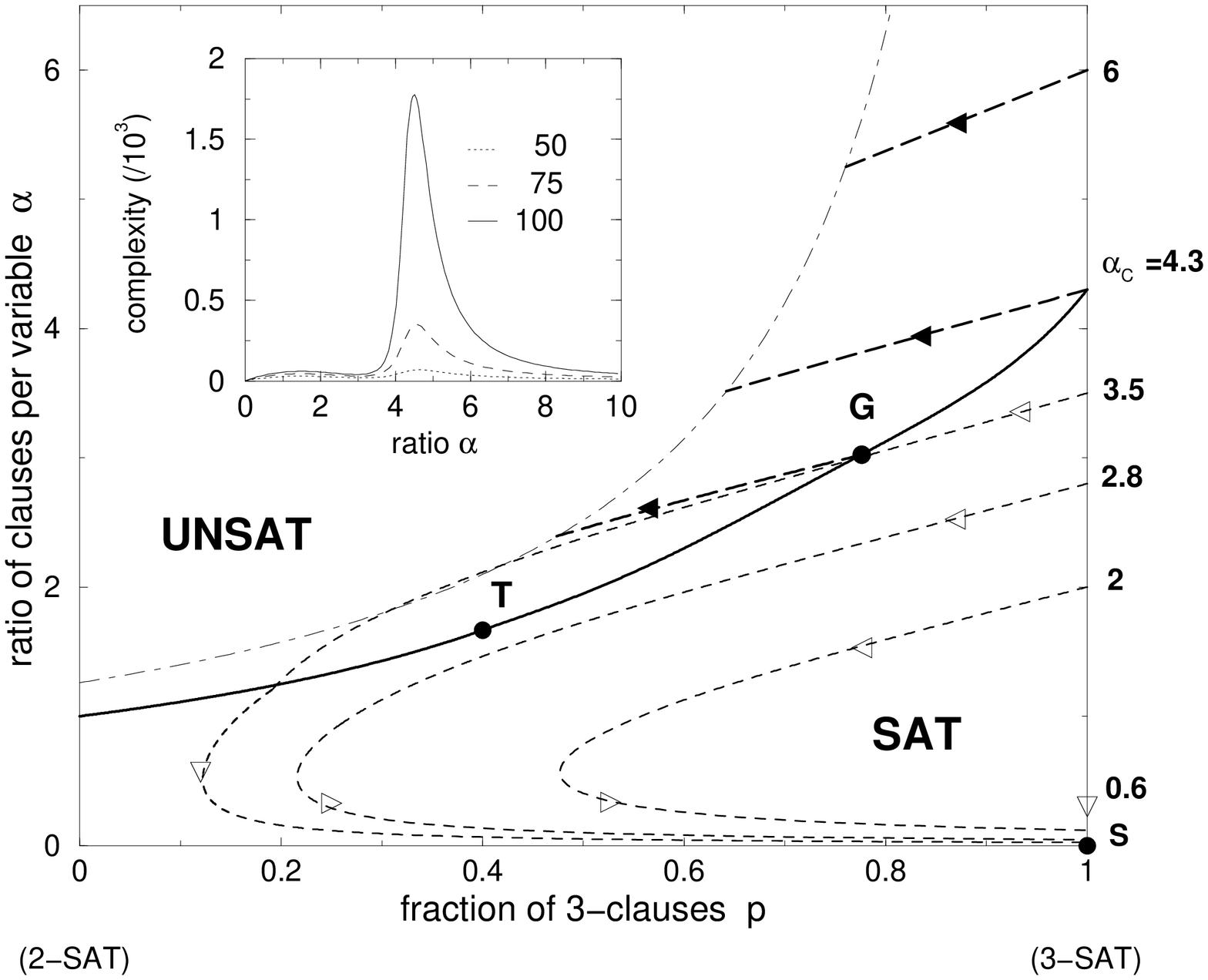}
\end{center}
\caption{Phase diagram of 2+p-SAT and resolution trajectories under
DPLL action. The threshold line $\alpha_C (p)$ (bold full line)
separates sat (lower part of the plane) from unsat (upper part)
phases.  Departure points for DPLL trajectories are located on the
3-SAT vertical axis.  Arrows indicate the direction of "motion" along
trajectories (dashed curves) parameterized by the fraction $t$ of
variables set by DPLL.  For small ratios $\alpha < \alpha _L$ ($\simeq
3.003$ for the GUC heuristic), branch trajectories remain confined to
the sat phase, end in $S$ of coordinates $(1,0)$, where a solution is
found (with a search process reported on Fig.~\ref{trees}A).  For
$\alpha > \alpha _C \simeq 4.3$, proofs of unsatisfiability are given
by complete search trees with all leaves carrying contradictions
(Fig.~\ref{trees}B).  The corresponding tree trajectories are
represented by bold dashed lines (full arrows), which end up on the
halting (dot-dashed) line, see text.  For ratios $\alpha _L < \alpha <
\alpha_C$, the branch trajectory intersects the threshold line at some
point $G$. A contradiction a.s. arises, and extensive backtracking up
to $G$ permits to find a solution (Fig.~\ref{trees}C).  With
exponentially small probability, the search tree looks like
Fig.~\ref{trees}A instead: the trajectory (dashed curve) crosses 
the "dangerous" region where contradictions are likely to occur,
then exits from this region and ends up with a solution (lowest dashed
trajectory). Inset: Resolution time of 3-SAT instances as a function of
the ratio of clauses per variable $\alpha$ and for three different
sizes.  Data correspond
to the median resolution time of 10,000 instances by DPLL; the average time
may be somewhat larger due to the presence of rare, exceptionally
hard instances, cf. Sec.~\ref{DpllFlucSection}. 
The computational complexity is linear for $\alpha < \alpha _L \simeq 3.003$,
exponential above. }
\label{sche}
\end{figure}

The input distribution of 3-SAT is characterized by a single parameter
$\pi$, the ratio $\alpha$ of clauses per variable.
The action of DPLL on an instance of 3-SAT, illustrated 
in Fig.~\ref{algo}, causes 
the changes of the overall numbers of variables and clauses, and thus 
of $\alpha$. Furthermore, DPLL reduces some 3-clauses to 2-clauses. We
use a mixed 2+p-SAT distribution\cite{Sta}, where $p (=\pi')$ is the fraction
of 3-clauses, to model what remains of the input instance at a node of
the search tree. Using experiments and methods from statistical
mechanics\cite{Sta} and rigorous calculations\cite{Achl1}, 
the threshold line $\alpha _C (p)$, separating
sat from unsat phases, may be estimated with the results shown in
Fig.~\ref{sche}. For $p \le p_0 = 2/5$, {\em i.e.} left to point T, 
the threshold line is
given by $\alpha _C(p)=1/(1-p)$, and saturates the upper bound
for the satisfaction of 2-clauses. Above $p_0$, no exact
value for $\alpha _C (p)$ is known.
The phase diagram of 2+p-SAT is the natural space in which the DPLL
dynamics takes place. An input 3-SAT instance with ratio $\alpha$ shows
up on the right vertical boundary of Fig.~\ref{sche} as a point of
coordinates $(p=1,\alpha )$.  Under the action of DPLL, the
representative point moves aside from the 3-SAT axis and follows a
trajectory in the $(p,\alpha )$ plane.

In this section, we show that the location of this trajectory in
the phase diagram allows a precise understanding of the search tree
structure and of complexity as a function of the ratio $\alpha$
of the instance to be solved (Inset of Fig.~\ref{sche}).  
In addition, we shall present an
approximate calculation of trajectories accounting for the case of
massive backtracking, that is for unsat instances, and slightly below
the threshold in the sat phase.  Our approach is based on a non
rigorous extension of works by Chao and Franco who first studied the
action of DPLL (without backtracking) on easy, sat
instances\cite{fra2,Fra} as a way to obtain lower bounds to the
threshold $\alpha_C$, see \cite{Achl} for a recent review.

Let us emphasize that the idea of trajectory is made possible thanks to an
important statistical property of the heuristics of split we 
consider \cite{fra2,Fra},
\begin{itemize}
\item{Unit-Clause (UC) heuristic:} 
pick up randomly a literal among a unit clause 
if any, or any unset variable otherwise.

\item{Generalized Unit-Clause (GUC) heuristic:}  
pick up randomly a literal  among the shortest avalaible clauses.

\item{Short Clause With Majority (SC$_1$) heuristic:}  
pick up randomly a literal among unit clauses if any, or pick up randomly 
an unset variable $v$, count
the numbers of occurences $\ell, \bar \ell$ of $v$, $\bar v$ in 3-clauses,
and choose $v$ (respectively $\bar v$) if $\ell > \bar \ell$ (resp.
$\ell < \bar \ell$). When 
$\ell=\bar \ell$, $v$ and $\bar v$ are equally likely to be chosen.
\end{itemize}

These heuristics do not induce any bias nor correlation in the
instances distribution\cite{fra2,kir2}. Such a statistical
``invariance'' is required to ensure that the dynamical evolution
generated by DPLL remains confined to the phase diagram of
Fig.~\ref{sche}.  In the following, the initial ratio of clauses per
variable of the instance to be solved will be denoted by $\alpha _0$.

\subsubsection{Lower sat phase and branch trajectories.}

Let us consider the first descent of the algorithm, that is the action
of DPLL in the absence of backtracking. The search tree is a single
branch (Fig.~\ref{trees}A). The numbers of 2 and 3-clauses 
are initially equal to $C_2=0, C_3=\alpha _0
\, N$ respectively. Under the action of DPLL, $C_2$ and $C_3$ follow 
a Markovian stochastic evolution process, as the depth $T$ along the branch 
(number of assigned variables) increases. Both $C_2$ and $C_3$ are 
concentrated around their average values, the densities 
$c_j (t)= E[C_j( t N)/N]$ ($j=2,3$) of which obey a set of 
coupled ordinary differential equations (ODE)\cite{fra2,Fra,Achl},
\begin{equation}
\frac{d c_3}{dt} = - \frac{ 3\, c_3}{1-t} \qquad , \qquad
\frac{d c_2}{dt} = \frac{ 3\, c_3}{2(1-t)} - \frac{ 2\, c_2}{1-t} -
\rho _1 (t) \; h(t) \qquad , \label{ode}
\end{equation}
where $\rho _1 (t) = 1 - c_2(t)/(1-t)$ is the probability that DPLL fixes a 
variable  at depth $t$ through unit-propagation. Function $h$ depends upon the
heuristic: $h_{UC} (t)=0$, $h_{GUC} (t)=1$ (if $\alpha _0> 2/3$), 
$h_{SC_1}
(t)=a\, e^{-a}\, (I_0(a)+I_1(a))/2$  with $a\equiv 3\, c_3(t)/(1-t)$
and $I_\ell$ denotes the $\ell^{th}$ modified Bessel function.
To obtain the single branch trajectory in the phase diagram of Fig.~\ref{sche},
we solve the ODEs (\ref{ode}) with initial conditions $c_2(0)=0, 
c_3(0)=\alpha_0$, and perform the change of variables
\begin{equation}
p(t) = \frac{c_3(t)}{c_2(t)+c_3(t)} \qquad , \qquad
\alpha (t) = \frac{c_2(t)+c_3(t)}{1-t} \qquad . \label{change}
\end{equation}

Results are shown for the GUC heuristics and starting ratios $\alpha_0 =2$
and 2.8 in Fig.~\ref{sche}. Trajectories,
indicated by light dashed lines, first head to the left and then
reverse to the right until reaching a point on the 3-SAT axis at
a small ratio. Further action of
DPLL leads to a rapid elimination of the remaining clauses and the
trajectory ends up at the right lower corner S, where a solution is
found.

Frieze and Suen \cite{Fri} 
have shown that, for ratios $\alpha _0 < \alpha _L \simeq 3.003$
(for the GUC heuristics), the full search tree essentially reduces
to a single branch, and is thus entirely described by the ODEs (\ref{ode}).
The number of backtrackings necessary to reach a solution 
is bounded from above 
by a power of $\log N$. The average size $\bar Q$ of the branch then 
scales linearly
with $N$ with a multiplicative factor $\gamma (\alpha _0)=Q/N$ that can
be analytically computed \cite{Coc}.

The boundary $\alpha _L$ of this easy sat region can be defined as the largest
initial ratio $\alpha _0$ such that the branch trajectory $p(t),\alpha (t)$ 
issued from $\alpha _0$ never leaves the sat phase in the course of DPLL 
resolution.

\subsubsection{Unsat phase and tree trajectories.}

For ratios above threshold ($\alpha _0 > \alpha _C\simeq 4.3$), instances
almost never have a solution but a considerable amount of
backtracking is necessary before proving that clauses are
incompatible. As shown in Fig.~\ref{trees}B, a generic unsat tree includes
many branches. The number of branches (leaves), $B$, or the number
of nodes, $Q=B-1$, grow exponentially with $N$\cite{Chv}.
It is convenient to define its logarithm $\omega$ through $B=2^{N
\omega}$. 
Contrary to the previous section, the sequence of points $(p,\alpha)$ 
characterizing the evolution of the 2+p-SAT instance solved by DPLL does not
define a line any longer, but rather a patch, or cloud of points with
a finite extension in the phase diagram of Fig.~\ref{schemoins}. 

\begin{figure}
\begin{center}
\includegraphics[height=150pt,angle=0]{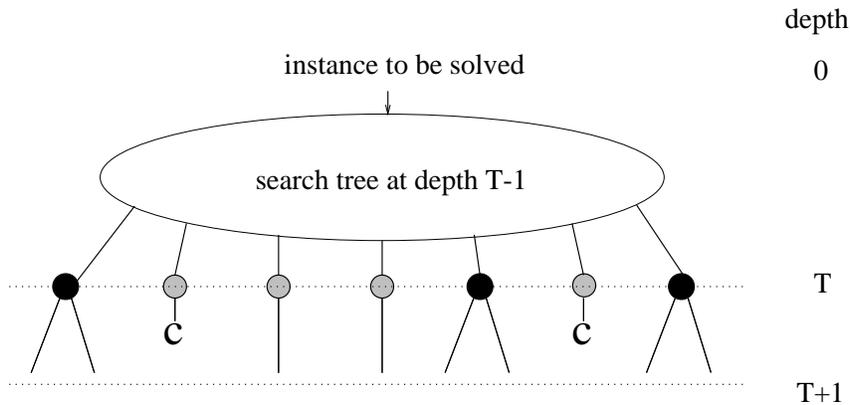}
\end{center}
\caption{Imaginary, parallel growth process of an unsat search tree used in the
theoretical analysis of the computational complexity. Variables 
are fixed through unit-propagation, or the splitting heuristics as in the DPLL 
procedure, but branches evolve in parallel. $T$ denotes the depth in the
tree, that is the number of variables assigned by DPLL along each (living) 
branch. At depth $T$, one literal is chosen on each branch among 1-clauses
(unit-propagation, grey circles not represented on Figure 1), or 2,3-clauses 
(split, black circles as in Figure 1).
If a contradiction occurs as a result of unit-propagation, the branch gets 
marked with C and dies out. The growth of the tree proceeds  
until all branches carry C leaves. The resulting tree is identical to the one
built through the usual, sequential operation of DPLL. }
\label{struct}
\end{figure}

We have analytically computed the logarithm $\omega$ of the size 
of these patches, as a function of $\alpha_0$,
extending to the unsat region the probabilistic analysis of DPLL. This
is, {\em a priori}, a very difficult task since the 
search tree of Fig.~1B is the output of a complex, sequential process: nodes
and edges are added by DPLL through successive descents and
backtrackings.  We have imagined a different building up, that results
in the same complete tree but can be mathematically analyzed: the tree
grows in parallel, layer after layer (Fig.~\ref{struct}).

\begin{center}
\begin{figure}
\includegraphics[height=160pt,angle=-90]{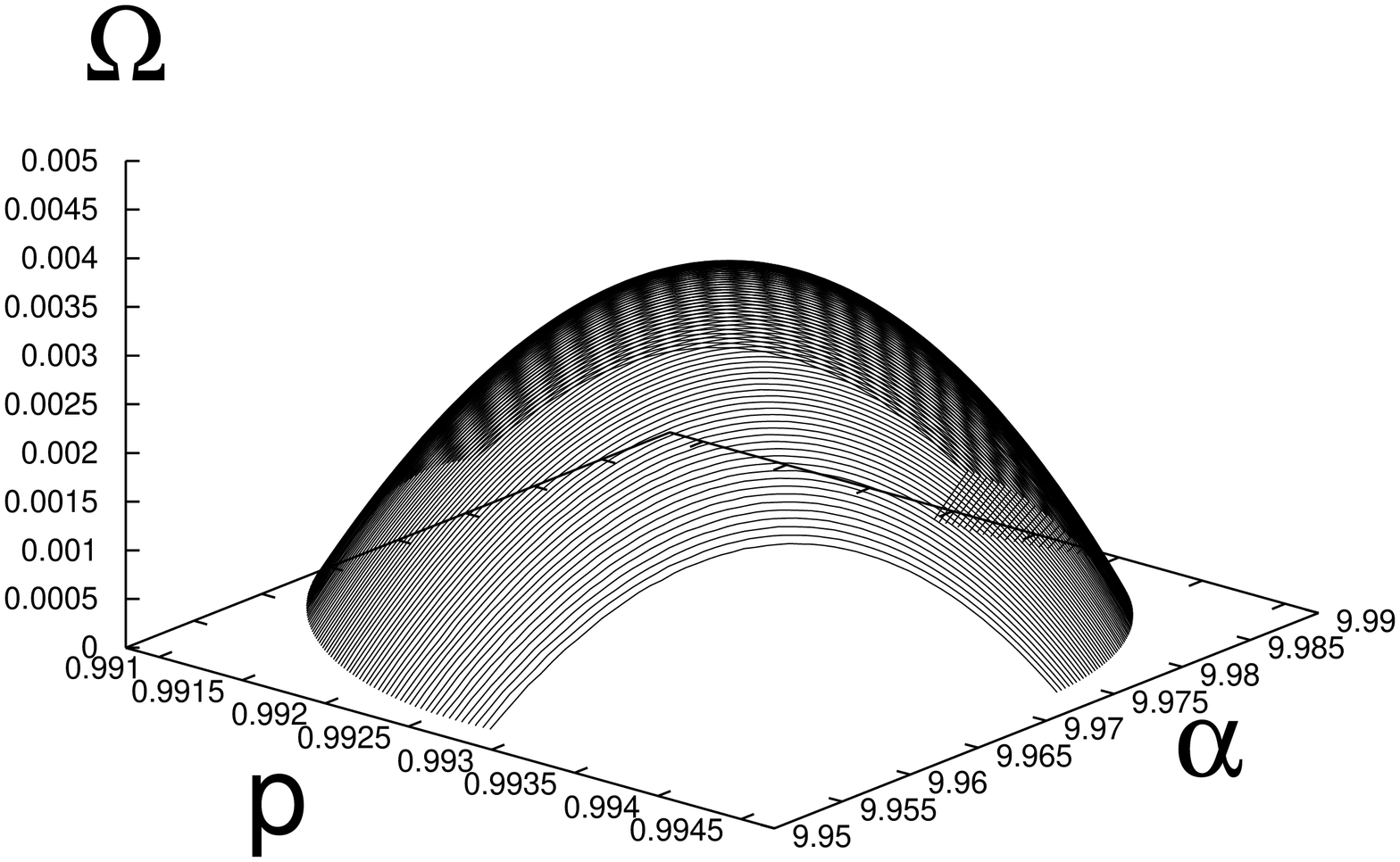}
\includegraphics[height=160pt,angle=-90]{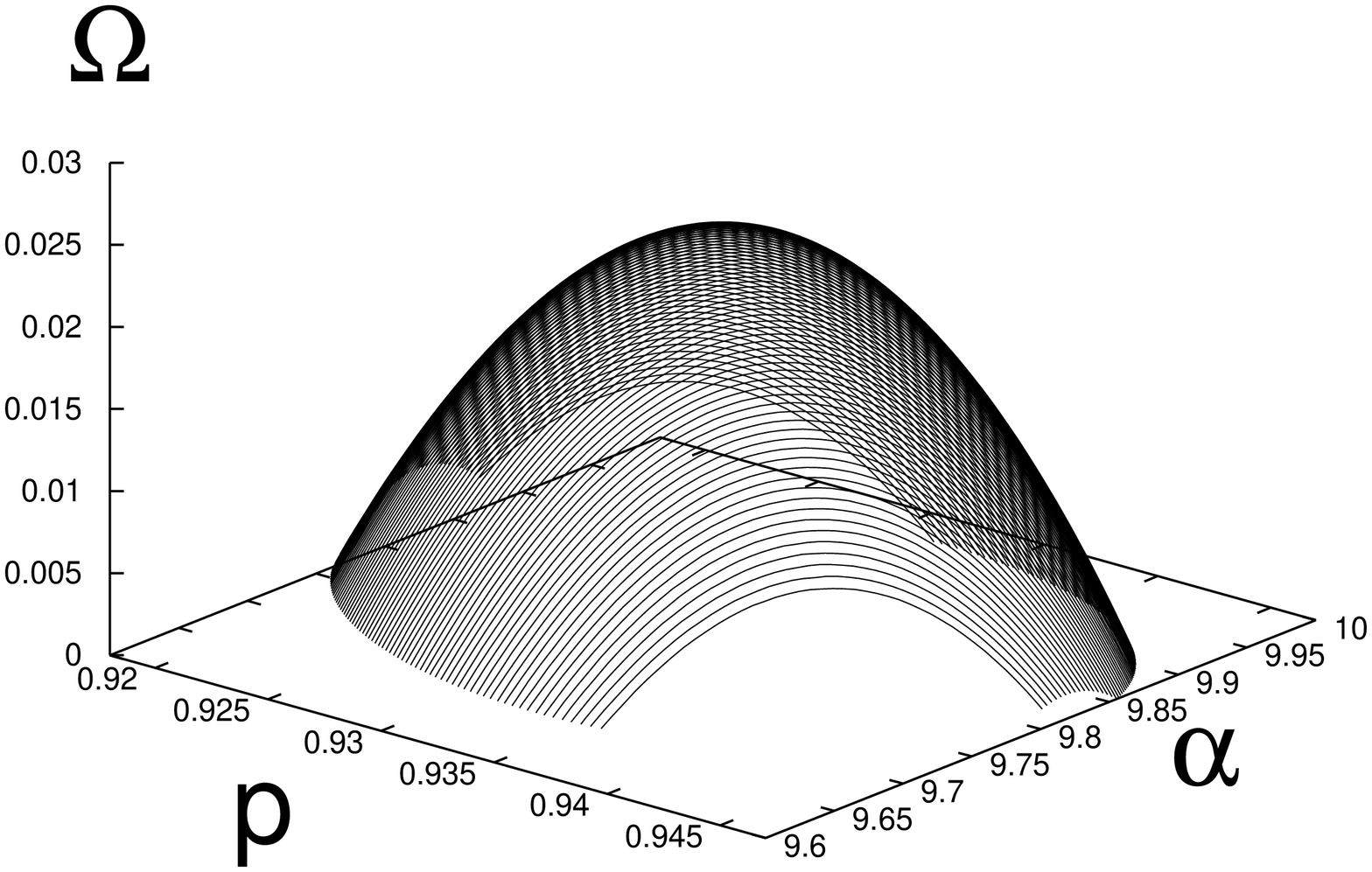} 
\includegraphics[height=160pt,angle=-90]{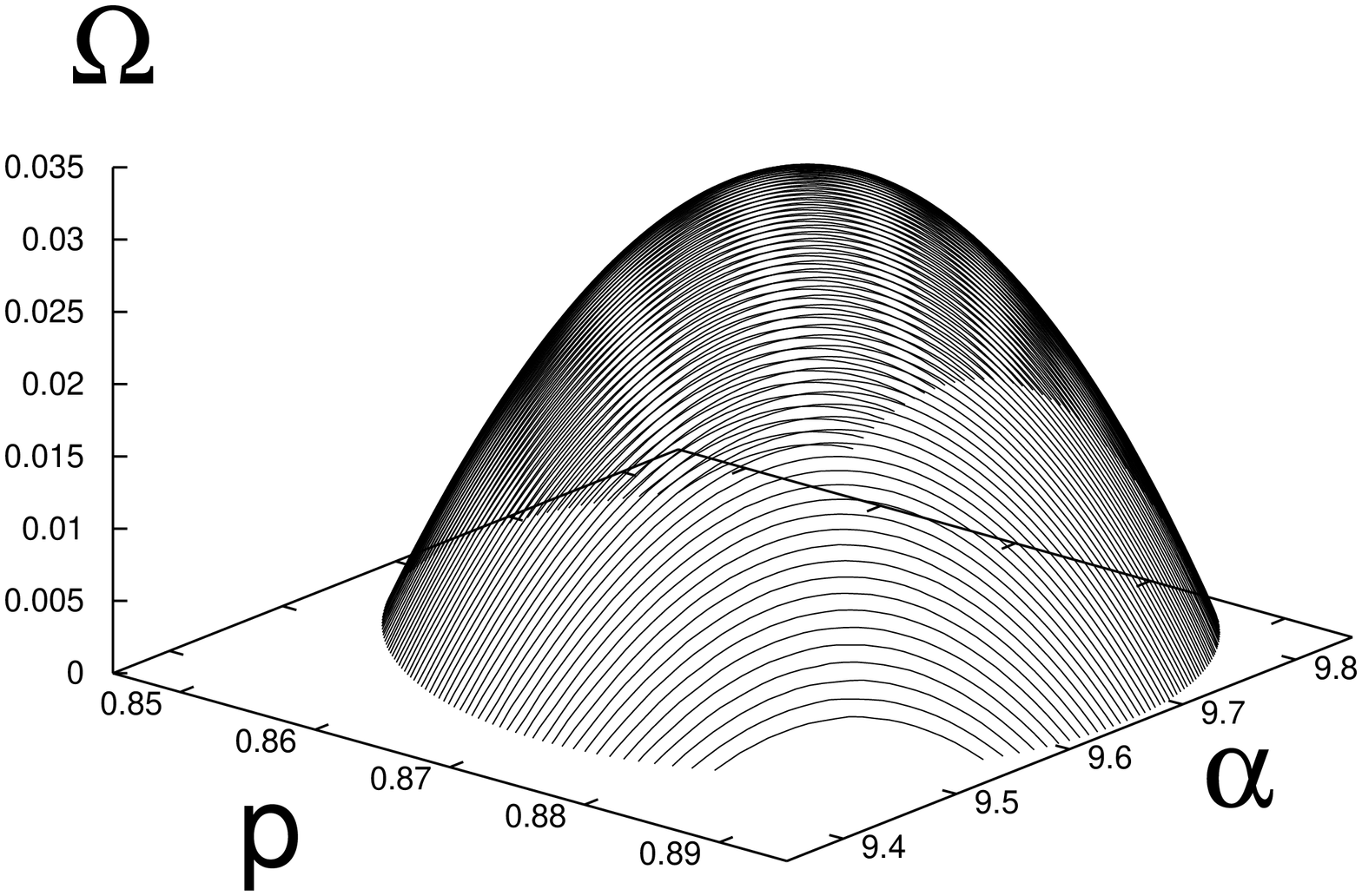}
\vskip .5cm
\caption{Snapshots of the surface $\omega (p,\alpha )$ for $\alpha
_0=10$ at three different depths, $t=0.01$, 0.05 and 0.09 (from left to
right). The height $\omega ^*(t)$ of the top of the surface, with
coordinates $p^*(t), \alpha^*(t)$, is the logarithm (divided by $N$) of the
number of branches. The halt line is hit at $t_h \simeq 0.094$.}
\label{dome}
\end{figure}
\end{center}

A new layer is added by assigning, according to DPLL heuristic, one
more variable along each living branch. As a result, a branch may
split (case 1), keep growing (case 2) or carry a contradiction and die
out (case 3).  Cases 1,2 and 3 are stochastic events, the
probabilities of which depend on the characteristic parameters
$c_2,c_3,t$ defining the 2+p-SAT instance carried by the branch, and
on the depth (fraction of assigned variables) $t$ in the tree.  We
have taken into account the correlations between the parameters
$c_2,c_3$ on each of the two branches issued from splitting (case 1),
but have neglected any further correlation which appear between
different branches at different levels in the tree\cite{Coc}.  This
Markovian approximation permits to write an evolution equation for the
logarithm $\omega(c_2,c_3,t)$ of the average number of branches with
parameters $c_2,c_3$ as the depth $t$ increases,

\begin{equation}
\frac{\partial \omega } {\partial t} (c_2,c_3,t) = { H} \left[ c_2, c_3,
\frac{\partial \omega } {\partial c_2} , \frac{\partial \omega }
{\partial c_3 } ,t \right] \qquad . \label{croi}
\end{equation}
${H}$ incorporates the details of the splitting
heuristics. In terms of the partial
derivatives $y_2=\partial \omega/ \partial c_2$, $y_3=\partial \omega/
\partial c_3$, we have for the UC and GUC heuristics 
\begin{eqnarray}
{H} _{UC} 
&=& 1 + \frac{1}{\ln2} \left[ \frac {3\, c_3}{1-t}\; \left( e^{y_3}
\frac{1+e^{-y_2}}{2} -1 \right)+ \frac{c_2}{1-t}  \; \left(  \frac 32 e^{-y_2}
-2 \right) \right] \nonumber \\
{H} _{GUC} 
&=& \log _2 \nu (y_2)  
+  \frac{1}{\ln2} \left[ \frac {3\, c_3}{1-t}\; \left( e^{y_3}
\frac{1+e^{-y_2}}{2} -1 \right)+ \frac{c_2}{1-t}  \; \left( \nu(y_2)
-2 \right) \right] \nonumber \\ \hbox{\rm where} &&
\nu (y_2 ) = \frac 12\; e^{y_2}\left(  1 +\sqrt{1+ 4 e^{-y_2}} \right)\qquad .
\end{eqnarray}
Partial differential equation (PDE) (\ref{croi}) is
analogous to growth processes encountered in statistical physics
\cite{Gro}.  The surface $\omega$, growing with ``time'' $t$ above the
plane $(c_2,c_3)$, or equivalently from (\ref{change}), above the plane 
$(p,\alpha)$ (Fig.~\ref{dome}), describes the whole distribution of branches.
The average number  of branches at depth $t$ in the tree equals
$B(t) = \int dp\; d\alpha \; 2^{N\, \omega (p,\alpha,t)} \simeq
2^{N\, \omega ^*(t)}$, 
where $\omega ^*(t)$ is the maximum over $p,\alpha$ of $\omega (p,\alpha,t)$
reached in $p^*(t), \alpha^*(t)$.
In other words, the exponentially dominant contribution to $B(t)$
comes from branches carrying 2+p-SAT instances with parameters
$p^*(t), \alpha^*(t)$, which define the tree trajectories on
Fig.~\ref{sche}. 

The hyperbolic line in Fig.~\ref{sche} indicates the halt points, where
contradictions prevent dominant branches from further growing.
Each time DPLL assigns a variable through
unit-propagation, an average number $u(p,\alpha)$ of new 1-clauses is
produced, resulting in a net rate of $u-1$ additional 1-clauses.
As long as $u< 1$, 1-clauses are quickly eliminated and do not
accumulate.  Conversely, if $u  >1$, 1-clauses tend to accumulate. 
Opposite 1-clauses $x$ and $\bar x$ are likely to appear,
leading to a contradiction \cite{Fra,Fri}. The halt line is defined through
$u (p,\alpha)=1$.  As far as dominant branches are concerned,
the equation of the halt line reads
\begin{equation}
\alpha = \left( \frac{3+\sqrt 5}2 \right) 
\ln \left[ \frac{1+\sqrt 5}2 \right]\;\frac 1{1-p}
\simeq \frac{1.256}{1-p}\qquad .
\end{equation}

Along the tree trajectory, $\omega ^*(t)$ grows from 0, on the
right vertical axis, up to some final positive value, $\hat \omega$,
on the halt line. $\hat \omega $ is our theoretical prediction for the
logarithm of the complexity (divided by $N$).  Values of $\hat \omega
$ obtained for $4.3<\alpha_0<20$ by solving equation (\ref{croi})
compare very well with numerical results \cite{Coc}.

We
have plotted the surface $\omega$ above the $(p,\alpha)$ plane, with the results
shown in Fig.~\ref{dome}. 
It must be stressed that, though our calculation is not rigorous, it provides
a very good quantitative estimate of the complexity. Furthermore, 
complexity is found to scale asymptotically as
\begin{equation}
\hat \omega (\alpha _0) \sim \frac {3+\sqrt{5}}{(6 \,\ln 2) \;
\alpha _0}\; \left[ \ln \left( \frac{1+\sqrt 5}{2} \right) 
\right]^2 \simeq \frac{0.292}{\alpha _0} \qquad  (\alpha _0 \gg  \alpha _C ) .
\end{equation}
This result exhibits the expected scaling\cite{Bea}, and 
could indeed be exact. As $\alpha _0$ increases, search 
trees become smaller and smaller, and 
correlations between branches, weaker and weaker.

\subsubsection{Upper sat phase and mixed branch--tree trajectories.}

The interest of the trajectory approach proposed in this paper is best
seen in the upper sat phase, that is ratios $\alpha _0$ ranging from
$\alpha _L$ to $\alpha _C$. This intermediate region juxtaposes branch
and tree behaviors, see Fig.~\ref{trees}C. The branch trajectory starts
from the point $(p=1,\alpha _0)$ corresponding to the initial 3-SAT
instance and hits the critical line $\alpha_c(p)$ at some point G with
coordinates ($p_G,\alpha_G$) after $N\;t_G$ variables have been
assigned by DPLL (Fig.~\ref{sche}).  The algorithm then enters the unsat
phase and generates 2+p-SAT instances with no solution. A dense
subtree, that DPLL has to go through entirely, forms beyond G till the
halt line (Fig.~\ref{sche}).  The size of this subtree, $2^{N\,(1-t_G)\,\hat
\omega _G}$, can be analytically predicted from our theory. G is the
highest backtracking node in the tree (Fig.~\ref{trees}C) reached back by
DPLL, since nodes above G are located in the sat phase and carry
2+p-SAT instances with solutions. DPLL will eventually reach a
solution. The corresponding branch (rightmost path in Fig.~\ref{trees}C) is
highly non typical and does not contribute to the complexity, since
almost all branches in the search tree are described by the tree
trajectory issued from G (Fig.~\ref{sche}).  We have checked experimentally
this scenario for $\alpha _0=3.5$. The coordinates of the average
highest backtracking node, $(p_G\simeq 0.78, \alpha _G \simeq 3.02$),
coincide with the analytically computed intersection of the single
branch trajectory and the critical line $\alpha_c(p)$\cite{Coc}.  As
for complexity, experimental measures of $\omega$ from 3-SAT instances
at $\alpha _0= 3.5$, and of $\omega _G$ from 2+0.78-SAT instances at
$\alpha _G =3.02$, obey the expected identity $\omega = \omega _G \;
(1-t_G)$ and are in very good agreement with theory\cite{Coc}. 
Therefore, the structure of search trees for 3-SAT reflects
the existence of a critical line for 2+p-SAT instances.

\subsection{Average analysis of the vertex cover of random graphs}
\label{DpllVcSection}

We now consider the VC problem, where inputs are random graphs
drawn from the $G(N,p=c/N)$ ensemble\cite{Bollo}. In other words,
graphs have $N$ vertices and the probability that a pair of vertices
are linked through an edge is $c/N$, independently of other edges. 
 When the number $X=xN$ of covering marks is lowered, the model undergoes 
a COV/UNCOV transition at some critical density of covers 
$x_{\rm c}(c)$ for $N\to \infty$.
For $x>x_{\rm c}(c)$, vertex covers of size $Nx$
exist with probability one, for $x<x_{\rm c}(c)$ the available covering 
marks are not sufficient.
The statistical mechanics analysis of Ref. \cite{WeHa1} gave the result

\begin{eqnarray}
x_{\rm c}(c)= 1- \frac{2W(c)+W(c)^2}{2c}\, , \;\;\;\;\;\;\;\;
\mbox{for}\;\; c<e\, ,\label{Critical_VC}
\end{eqnarray}

where $W(c)$ solves the equation $We^W=c$.
This result is compatible with the bounds of Refs. \cite{Ga_VC,Fr_VC},
and was later shown to be exact \cite{Bauer}. 
For $c>e$, Eq.~(\ref{Critical_VC}) only gives an approximate estimate
of $x_{\rm c}(c)$. More sophisticated calculations can be found in Ref.
\cite{WeHaLong}.

\begin{figure}
\begin{center}
\includegraphics[height=220pt,angle=0] {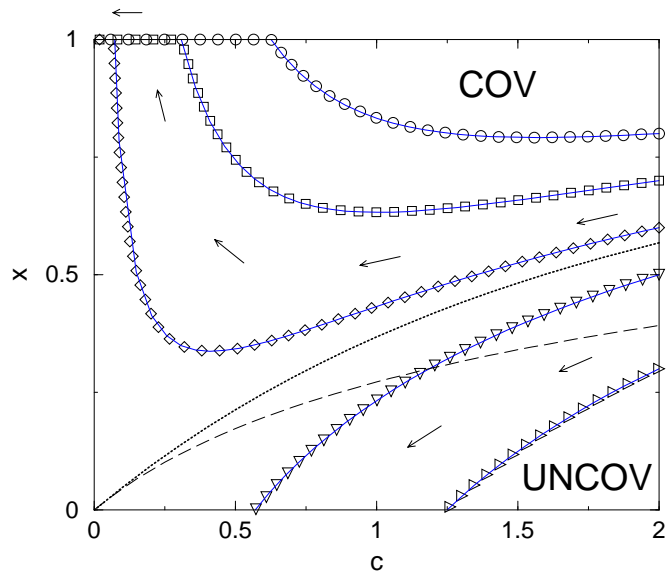}
\end{center}
\caption{Phase diagram of VC. The low-$x$, high-$c$ UNCOV phase is 
separated by the dashed line, cf. Eq. (\ref{Critical_VC}), from 
the high-$x$, low-$c$ COV phase. The symbols (numerics) and continuous lines
(analytical prediction, cf. Eq. (\ref{EqTrajVC})) 
refer to the simple search algorithm described in the text.
The dotted line is the separatrix between two types of 
trajectories.}
\label{traj_VC}
\end{figure}

Let us consider a simple implementation of the DPLL procedure for the present
problem.
During the computation,  vertices can be {\em covered}, 
{\em uncovered} or just {\em free}, 
meaning that the algorithm has not yet assigned any value to that
vertex. At the beginning all the vertices are set {\it free}. 
At each step the algorithm chooses a vertex $i$ at
random among those which are {\em free}. 
If $i$ has neighboring vertices which are either {\em free} or {\em
uncovered}, then the vertex $i$ is declared {\em covered} first. In case $i$
has only covered neighbors, the vertex is declared {\em uncovered}. The
process continues unless the number of covered vertices exceeds $X$.
In this case the algorithm backtracks and 
the opposite choice is taken for the vertex
$i$ unless this corresponds to declaring {\em uncovered} a vertex that
has one or more {\em uncovered} neighbors. 
The algorithm halts if it finds a solution
(and declares the graph to be COV) or after exploring all the search tree (in
this case it declares the graph to be UNCOV).

\begin{figure}
\begin{center}
\includegraphics[height=220pt,angle=0] {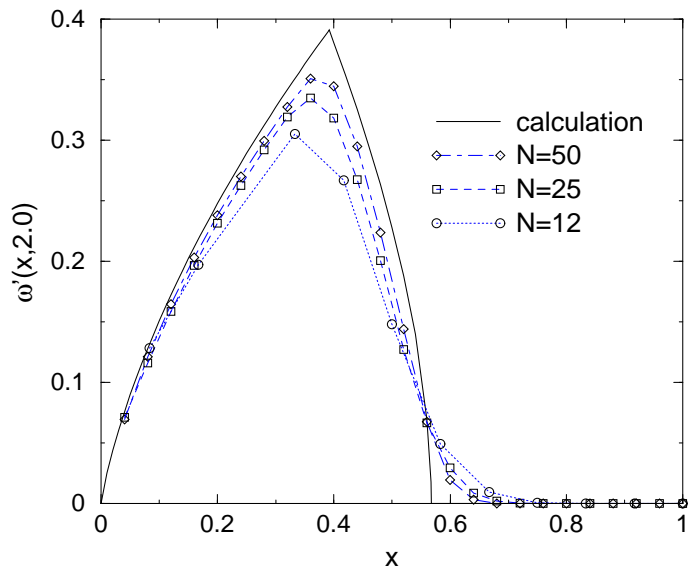}
\end{center}
\caption{Number of operations required to solve (or to show that no solution
exists to) the VC decision problem with the search algorithm described in 
the text. The logarithm of the number of nodes 
of the backtracking tree divided by the size $N$, is plotted versus the
number of covering marks. Here we consider random instances
with average connectivity $c=2$. The phase transition is at 
$x_{\rm c}(c=2)\approx 0.3919$ and corresponds to the peak in computational 
complexity.}
\label{time_VC}
\end{figure}

Of course one can improve over this algorithm by using smarter
heuristics \cite{WeHeur}. One remarkable example is the
``leaf-removal'' algorithm defined in Ref. \cite{Bauer}.
Instead of picking any vertex randomly, one chooses a connectivity-one 
vertex, declare it {\it uncovered}, and declare {\it covered} its neighbor.
This procedure is repeated iteratively on the subgraph
of {\it free} nodes, until no connectivity-one nodes are left.
In the low-connectivity, COV region $\{ c<e, x>x_{\rm c}(c)\}$,
it stops only when the graph is completely covered.
As a consequence, this algorithm can solve VC in linear time
with high probability in all this region. No equally good heuristics exists 
for higher connectivity, $c>e$.

\subsubsection{Branch trajectories}

Under the action of one of the above algorithms, the instance is progressively 
modified and the number of variables is reduced. 
In fact, at each step a vertex
is selected and can be eliminated from the graph regardless whether it is
declared {\it covered} or {\it uncovered}.
The analysis of the first algorithm is greatly simplified by the remark 
that, as long as backtracking has not begun, the new vertex is selected 
randomly. This implies that the modified instance 
produced by the algorithm is still a random graph.
Its evolution can be effectively described by a 
trajectory in the $(c,x)$ space. 
If one starts from  the parameters $c_0$, $x_0$, 
after $Nt$ steps of the algorithm, he will end up 
with a new instance of size $N(1-t)$ and parameters \cite{WeHa2}

\begin{eqnarray}
c(t)=c_0(1-t)\, , \;\;\;\;\; x(t)= 
\frac{x_0-t}{1-t}+\frac{e^{-c_0(1-t)}-e^{-c_0}}{c_0(1-t)}\, .\label{EqTrajVC}
\end{eqnarray}

Some examples of the two types of trajectories (the ones leading to a solution
and the ones which eventually enter the UNCOV region) are 
shown in Fig. \ref{traj_VC}. The separatrix is given by 

\begin{eqnarray}
x_{\rm s}(c) = 1-\frac{1-e^{-c}}{c}\, ,\label{Separatrix_VC}
\end{eqnarray}

and corresponds to the dotted line in Fig. \ref{traj_VC}. Above this line the
algorithm solves the problem in linear time.

For more general heuristics the analysis becomes less 
straightforward because the graph produced by the algorithm
does not belong to the standard random-graph {\it ensemble}. 
It may be necessary to augment the number of parameters which
describe the evolution of the instance. As an example,
the leaf-removal algorithm mentioned in the previous Section
is conveniently described by keeping track of three numbers
which parametrize the degree profile (i.e. the 
fraction of vertices $p_d(t)$ having a given degree $d$) of the 
graph \cite{WeHeur}.

\subsubsection{Tree trajectories}

Below the critical line $x_{\rm c}(c)$, cf. Eq. (\ref{Critical_VC}),
no solution exists to the typical random instance of VC. Our algorithm
must explore a large backtracking tree to prove it 
and this takes an exponential time.
The size of the backtracking tree could be computed along the lines
of Sec. II.B.2. However a good result can be obtained with a
simple ``static'' calculation \cite{WeHa1}. 

As explained in Sec. II.B.2, we imagine the evolution of the backtracking
tree as proceeding ``in parallel''.
At the level $M$ of the tree a set of $M$ vertices has been visited.
Call ${\cal G}_M$ the subgraph induced by these vertices.
Since we always put a covering mark on a vertex which is surrounded by 
vertices declared {\it uncovered}, each node 
of the backtracking tree will carry a vertex cover
of the associated subgraph ${\cal G}_M$ . Therefore the number of backtracking
nodes is given by

\begin{eqnarray}
Q = \sum_{M=1}^N {\cal N}_{\rm VC}({\cal G}_M;X)\, ,
\end{eqnarray}

where ${\cal N}_{\rm VC}({\cal G}_M;X)$ is the number of VC's of 
${\cal G}_M$ using at most $X$ marks. A very crude estimate of the 
right-hand side of the above equation is:

\begin{eqnarray}
Q \le \sum_{M=1}^N \sum_{X'=0}^{{\rm min}(X,M)}
\left(\begin{array}{c}M\\X'\end{array}\right)\, ,
\end{eqnarray}

where we bounded the number of VC's of size $X'$ on ${\cal G}_M$
with the number of ways of placing $X'$ marks on $M$ vertices.
The authors of \cite{WeHa2} provided a refined estimate based on 
the {\it annealed approximation} of statistical mechanics.
The results of this calculation are compared in Fig. \ref{time_VC}
with the numerics.

\subsubsection{Mixed trajectories}

If the parameters which characterize an instance of VC lie in the region 
between $x_{\rm c}(c)$, cf. Eq. (\ref{Critical_VC}),
and $x_{\rm s}(c)$, cf. Eq. (\ref{Separatrix_VC}), the problem is 
still soluble but our algorithm takes an exponential time to solve it.
In practice, after a certain number of vertices has been visited and 
declared either {\it covered} or {\it uncovered}, the remaining  subgraph
${\cal G}_{free}$ cannot be any longer covered with the leftover marks. 
This happens typically when the first descent trajectory (\ref{EqTrajVC})
crosses the critical line (\ref{Critical_VC}).

It takes some time for the algorithm to realize this fact. More precisely,
it takes exactly the time necessary to prove that ${\cal G}_{free}$
is uncoverable. This time dominates the computational complexity in 
this region and can be calculated along the lines sketched in the 
previous Section. The result is, once again, reported in Fig. \ref{time_VC},
which clearly shows a computational peak at the phase boundary. 

Finally, let us notice that this mixed behavior disappears in the
entire $c<e$ region if the leaf-removal heuristics is adopted for the 
first descent.

\subsection {Distribution of resolution times}
\label{DpllFlucSection}

\begin{figure}
\begin{center}
\includegraphics[height=220pt,angle=-90] {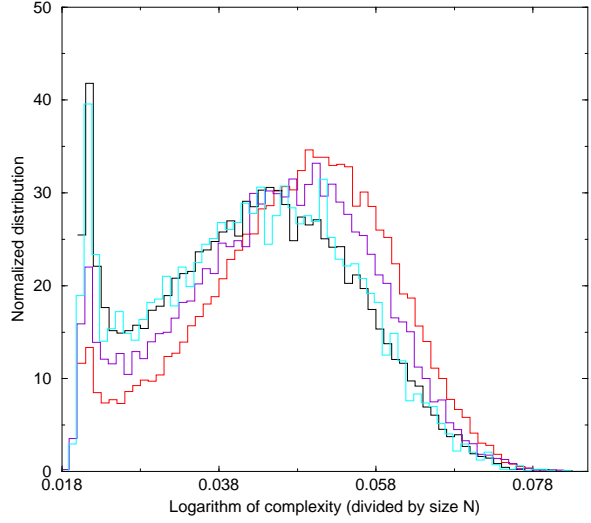}
\end{center}
\caption{Probability distributions of the logarithm $\omega$ of the 
resolution complexity from 20,000 runs of DPLL on random 3-SAT instances
with ratio $\alpha =3.5$. Each  
distribution corresponds 
to one randomly drawn instance of size $N=300$.}
\label{historun}
\end{figure}

\begin{figure}
\begin{center}
\includegraphics[height=220pt,angle=-90] {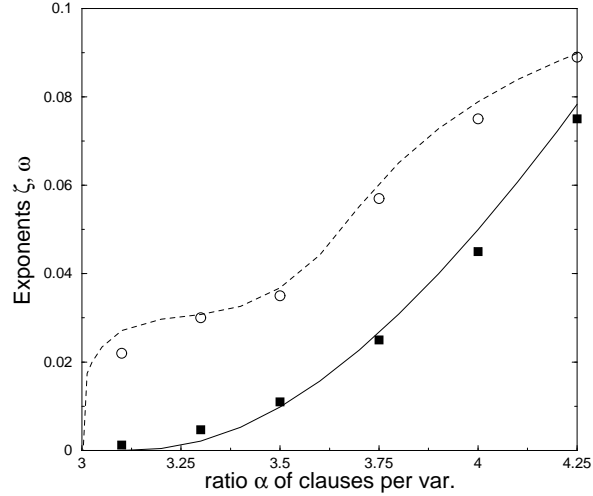}
\end{center}
\caption{Resolution of random 3-SAT instances in the upper sat phase: 
logarithm of complexity with DPLL ($\omega$ -- simulations:
circles, theory: dotted line) and restarts
($\zeta$ -- simulations: squares, theory: full line) 
as a function of ratio $\alpha$. Inset: Minus log. of 
the cumulative probability $P_{lin}$  of complexities 
$Q\le N$ as a function of the size for $100 \le N\le 400$ 
(full line); log. of the number of restarts $N_{rest}$ 
necessary to find a solution for $100\le N \le 1000$ 
(dotted line) for $\alpha=3.5$. Slopes are $\zeta = 0.0011$ and $\bar
\zeta = 0.00115$ respectively.}
\label{histolin}
\end{figure}

Up to now we have studied the typical resolution complexity. The study of
fluctuations of resolution times is interesting too, particularly in
the upper sat phase where solutions exist but are found at a price of
a large computational effort. We may expect that there exist lucky but
rare resolutions able to find a solution in a time much smaller 
than the typical one.
Due to the stochastic character  of DPLL 
complexity indeed fluctuates from run to run of the algorithm on the same
instance. In Fig.~\ref{historun} we show this run-to-run distribution of
the logarithm $\omega$ of the resolution complexity for four
instances of random 3-SAT with the same ratio $\alpha =3.5$. 
The run to run distribution are qualitatively independent of the
particular instances, and exhibit two bumps. The wide right one, 
located in $\omega \simeq 0.035$, correspond to the major part
of resolutions. It acquires more and more weight as $N$ increases and
corresponds to the typical behavior analysed in Section~II.B.3.
The left peak corresponds to much faster resolutions, taking place in
linear time. The weight of this peak (fraction of runs with complexities
falling in the peak) decreases exponentially fast with $N$, and can be
numerically estimated to $W_{lin} = 2^{- N \zeta}$
with $\zeta \simeq 0.011$.
Therefore, instances at $\alpha =3.5$ are typically solved in
exponential time but a tiny (exponentially small) fraction of runs 
are able to find a solution in linear time only. 

A systematic stop-and-restart procedure may be introduced to take
advantage of this fluctuation phenomenon and speed up resolution.  If
a solution is not found before $N$ splits, DPLL is stopped and rerun
after some random permutations of the variables and clauses.  The
expected number $N_{rest}$ of restarts necessary to find a solution
being equal to the inverse probability $1/W_{lin}$ of linear
resolutions, the resulting complexity scales as $N\; W_{lin}^{-1} \sim
2^{N\, \zeta}$.

To calculate $\zeta$ we have analyzed, along the lines of the
study of the growth of the search tree in the unsat phase, 
the whole distribution of the complexity 
for a given ratio $\alpha$ in the upper sat phase.
Calculations can be found in \cite{Cocrs}.
Linear resolutions are found to correspond to branch trajectories that 
cross the unsat phase without being hit by a contradiction,
see Fig.~\ref{sche}.
Results are reported in Fig.~\ref{histolin} and compare very well
with the experimentally measured number $N_{rest}$ 
of restarts necessary to find a solution.
In the whole upper sat phase, the use of restarts offers an exponential gain
with respect to usual DPLL resolution (see Fig.~\ref{histolin}
for comparison between $\zeta$ and $\omega$), but the completeness 
of DPLL is lost.

\begin{figure}
\centerline{\includegraphics[height=220pt,angle=-90] {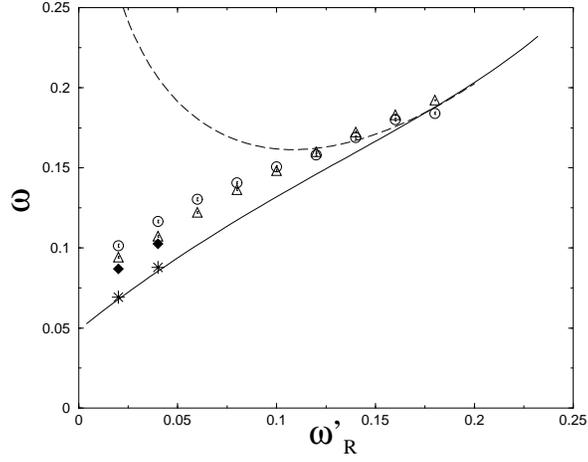}}
\caption{The computational complexity of the search 
algorithm for VC, with restarts after $\exp(N\omega'_R)$ 
backtracking steps. The complexity is defined as the 
logarithm of the total number of visited nodes, divided by the 
size $N$ of the graph. Symbols refer to $N=30$ (circles), $60$ (triangles),
and $120$ (diamonds).   The stars are the result
of an $N\to\infty$ extrapolation. 
The continuous (dashed) line reproduces the theoretical 
prediction with (without) taking into account fluctuations of the first 
descent trajectory.}
\label{time_RVC}
\end{figure}

A slightly more general restart strategy consists in stopping the backtracking
procedure after a fixed number of nodes $Q_R = e^{N\omega'_R}$ 
has been visited.
A new (and statistically independent) DPLL procedure is then 
started from the beginning. 
In this case one exploits lucky, but still 
exponential, stochastic runs. The tradeoff between
the exponential gain of time and the exponential number of restarts,
can be optimized by tuning the parameter 
$\omega'_R$. This approach has been analyzed in Ref. \cite{VCRestart} 
taking VC as a working example. 
In Fig. \ref{time_RVC} we show the computatonal complexity of
such a strategy as a function of the restart parameter $\omega'_R$. We
compare the numerics with an approximate calculation \cite{VCRestart}.
The instances were random graphs with average connectivity
$c=3.2$, and $x=0.6$ covering marks per vertex.
The optimal choice of the parameter seems to be (in this case)
$\omega'_R\approx 0$, corresponding to polynomial runs.

\begin{figure}
\begin{center}
\includegraphics[height=220pt,angle=-90] {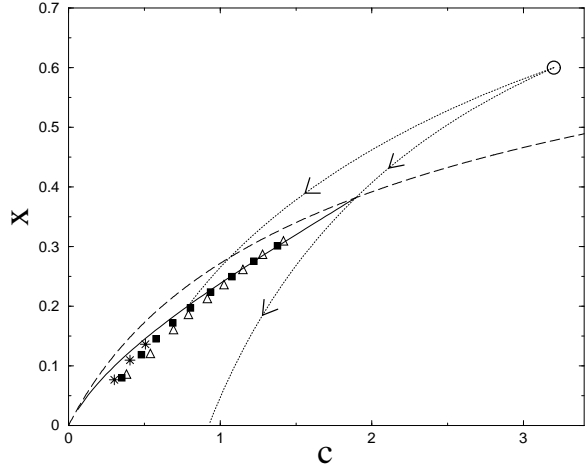}
\end{center}
\caption{Restart experiments for VC with initial condition at
$c_0=3.2$, $x_0=0.6$
(empty circle). The long-dashed line is the critical line
(\ref{Critical_VC}). The rightmost dotted line is the typical trajectory. 
The leftmost one is the rare trajectory followed by the last 
(successful) restart of the algorithm when $\omega'_R=0.1$. 
The symbols are numerical results for the
$(c,x)$ coordinates of the root of the backtrack tree generated by the
algorithm since the last restart. 
Triangles, squares and stars correspond,
respectively, to $N=30$, $60$, $120$ (in each case we considered
several values for $\omega'_R$, each one corresponding to a symbol). 
The continuous line is an approximate analytical
prediction for the same quantity.}
\label{root_RVC}
\end{figure}

The analytical prediction  reported in Fig. \ref{time_RVC}
requires, as for 3-SAT, an estimate of the execution-time fluctuations
of the DPLL procedure (without restart).
It turns out that one major source of fluctuations is, in the present case,
the location in the $(c,x)$ plane of the highest node 
in the backtracking tree. 
In the typical run this coincides with the intersection $(c_G,x_G)$ between 
the first descent trajectory (\ref{EqTrajVC})
and the critical line (\ref{Critical_VC}). One can estimate the probability
$P(c,x) \sim \exp\{-N\psi(c,x)\}$  for this node to have coordinates
$(c,x)$ (obviously $\psi(c_G,x_G) = 0$). 

When an upper bound $\omega'_R$ on the backtracking time is fixed,
the problem is solved in those lucky runs which are characterized
by an atypical highest backtracking node. Roughly speaking, this means
that the algorithm has made some very good (random) choices in its first 
steps. In Fig. \ref{root_RVC} we plot the position of the highest 
backtracking point in the (last) successful runs for several values of 
$\omega'_R$. Once again the numerics compare favourably with 
an approximate calculation.

\section{Analysis of local search algorithms}
\label{LocalSection}

We now turn to the description and study of algorithms of another
type, namely local search algorithms. As a common feature, these
algorithms  start from a configuration (assignment) of the variables,
and then make successive improvements by changing at each step few
of the variables in the configuration (local move).
For instance, in the SAT
problem, one variable is flipped from being true to false, or {\em vice
versa}, at each step. Whereas complete algorithms of the DPLL type
give a definitive answer to any instance of a decision problem,
exhibiting either a solution or a proof of unsatisfiability, local
search algorithms give a sure answer when a solution is found but
cannot prove unsatisfiability. However, these algorithms can sometimes be
turned into one-sided probabilistic algorithms, with an 
upper bound on the probability that a solution exists and has not
been found after $T$ steps of the algorithm, decreasing to zero
when $T\to\infty$\cite{rando}.

\subsection{Landscape and search dynamics}

\begin{figure}
\centerline{
\includegraphics[scale=0.3,angle=0]{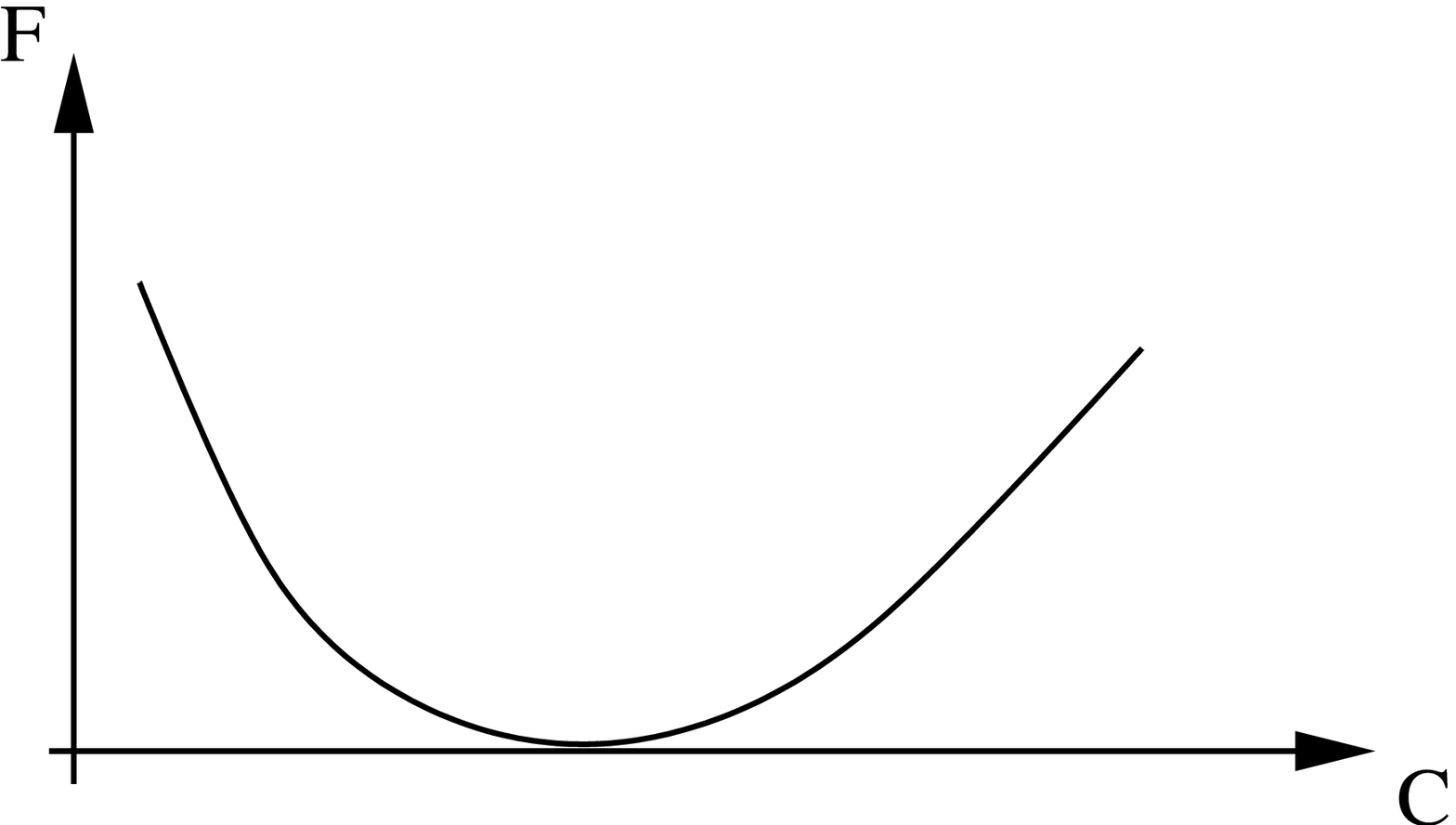}
\includegraphics[scale=0.3,angle=0]{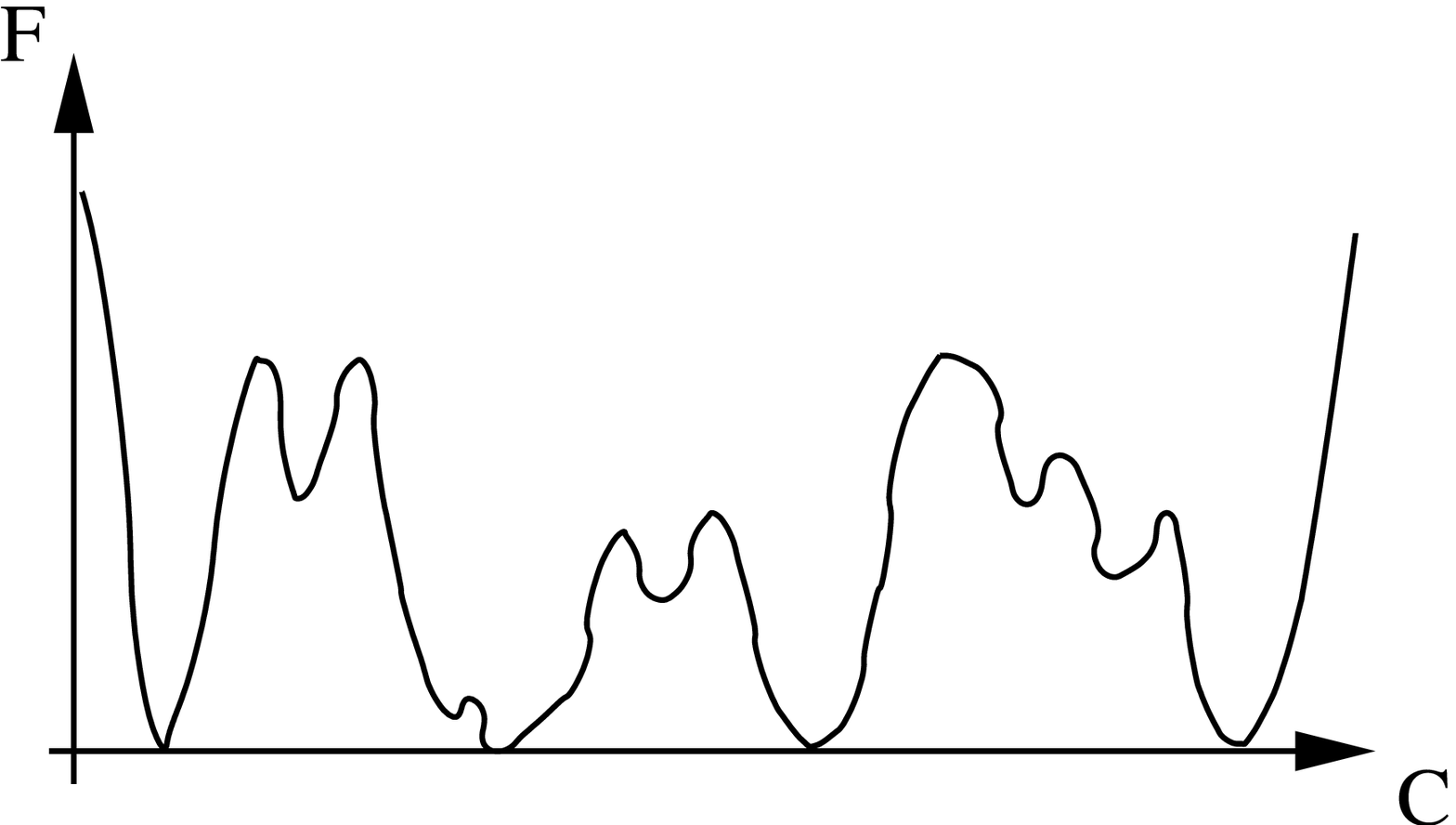}
\includegraphics[scale=0.3,angle=0]{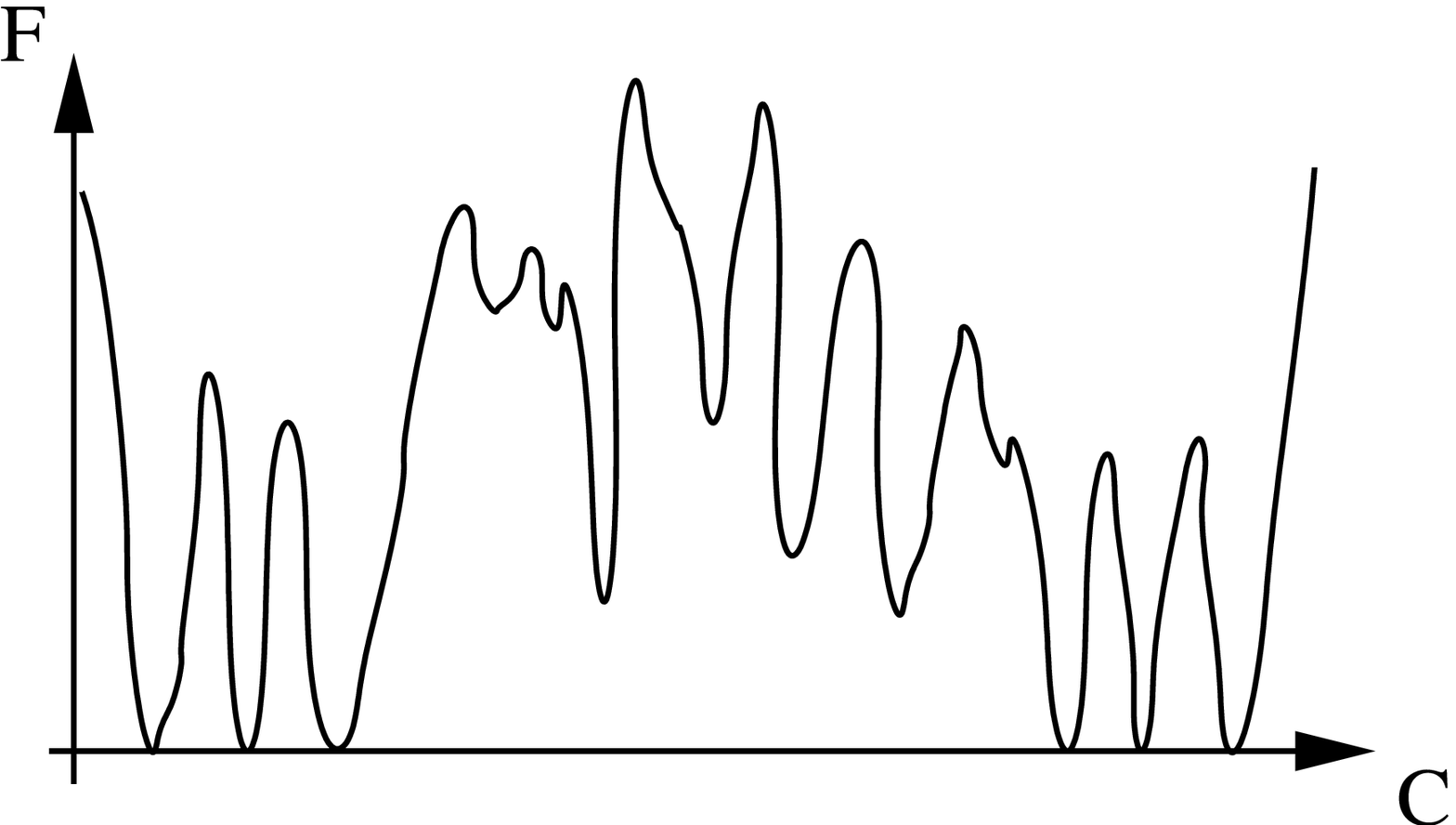}
}
\caption{Landscapes corresponding to three different cost functions.
Horizontal axis represent the space of configurations $C$, while vertical
axis is the associated cost $F(C)$. Left: smooth cost function,
with a single minimum easily reachable with local search procedures
e.g. gradient descent. Middle: rough cost function with a lot of 
local minima whose presence may damage the performances of local search 
algorithms. The various global minima are spread out homogeneously
over the configuration space. Right: rough cost function with 
global minima clustered in some portions of the configuration space only. }
\label{landscape}
\end{figure}

Local search algorithms perform repeated changes of a configuration $C$ 
of variables (values of the Boolean variables for SAT, status --marked 
or unmarked-- of vertices for VC) according to some criterion, usually 
based on the comparison of the cost function $F$ (number of 
unsatisfied clauses for SAT, of uncovered
edges for VC) evaluated at $C$ and over its neighborhood. 
It is therefore clear that the shape of the multidimensional surface 
$C \to F(C)$, called cost function landscape, is of high importance.
On intuitive grounds, if this landscape is relatively smooth with a
unique minimum, local procedures as gradient descent should be very
efficient, while the presence of many local minima could hinder the
search process (Fig.~\ref{landscape}). The fundamental underlying
question is whether the performances of the
dynamical process (ability to find the global minimum, time needed to
reach it) can be understood in terms of an analysis of the
cost function landscape only. 

This question was intensively studied and answered for a limited class
of cost functions, called mean field spin glass models, some years 
ago\cite{leti}.
The characterization of landscapes is indeed of huge importance in
physical systems. There, the cost function is simply the physical
energy, and local dynamics are usually low or zero temperature Monte
Carlo dynamics, essentially equivalent to gradient descent.
Depending on the parameters of the input distribution, the minima 
of the cost functions may undergo structural changes, a phenomenon
called clustering in physics. 

Clustering has been rigorously shown to take place in the random 3-XORSAT
problem\cite{Crei,xorsat1,xorsat2}, 
and is likely to exist in many other random combinatorial
problems as 3-SAT\cite{varia,cavi}. Instances of the 3-XORSAT problem with 
$M=\alpha\,N$ clauses and $N$ variables have almost surely
solutions as long as $\alpha < \alpha_c \simeq 0.918$\cite{xorsat1,xorsat2}. 
The clustering
phase transition takes place at $\alpha_s \simeq 0.818$ and is related
to a change in the geometric structure of the space of solutions,
see Fig.~\ref{landscape}:
\begin{itemize}
\item when $\alpha < \alpha_s$, the space of solutions is connected. Given
a pair of solutions $C,C'$, {\em i.e.} two assignments of the $N$ 
Boolean variables that satisfy the clauses, there almost surely exists 
a sequence of solutions, $C_j, j=0,1,2,\ldots, J$, with $C_0\equiv C$,
$C_J\equiv C'$, $J=O(N)$, 
connecting the two solutions such that the Hamming distance
(number of different variables) between $C_j$ and $C_{j+1}$ is bounded
from above by some finite constant when $N\to \infty$.
\item when $\alpha_s<\alpha<\alpha_c$, 
the space of solutions is not connected any longer.
It is made of an exponential (in $N$) number of connected components,
called clusters, each containing an exponentially large number of
solutions. Clusters are separated by large voids: the Hamming distance
between two clusters, that is, the smallest Hammming distance between
pairs of solutions belonging to these clusters, is of the order of $N$.
\end{itemize}
From intuitive grounds,
changes of the statistical properties of the cost function landscape e.g. 
of the structure of the solutions space may potentially affect
the search dynamics. This connection between dynamics and static properties
was established in numerous works in the context of 
mean field models of spin glasses \cite{leti}, and 
subsequently also put forward in some studies of local search algorithms in
combinatorial optimization problems\cite{varia,suedois,cavi}. 
So far, there is no satisfying explanation to when and why features of 
{\em a priori} algorithm dependent dynamical phenomena 
should be related to, or predictable from  
some statistical properties of the cost function landscape.
We shall see some examples in the following where such a connection
indeed exist (Sec.~III.B) and other ones where its presence is far less 
obvious (Sec.~IIIC,D).

\subsection{Algorithms for error correcting codes}
\label{CodeSection}

\def\utx{\underline{\tx}}
\def\tx{{\tt x}}
\def\utz{\underline{\tz}}
\def\tz{{\tt z}}
\def\ut0{\underline{\t0}}
\def\t0{{\tt 0}}

\begin{figure}
\centerline{
\includegraphics[height=220pt,angle=0] {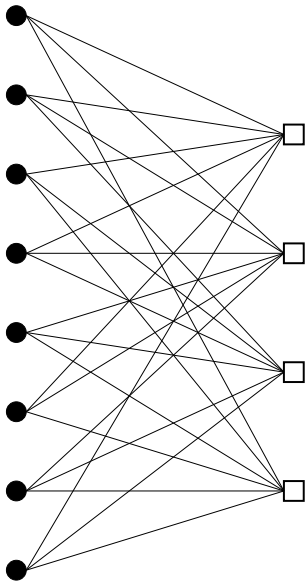}
}
\caption{Tanner graph of a {\it regular}
linear code. A left-hand node is associated to each
  variable, and a right hand node to each parity check.
A link is drawn between two nodes whenever the variable associated to
the left-hand one enters in the parity check corresponding to the right-hand 
one.}
\label{Tanner}
\end{figure}

Coding theory is a rich source of computational problems (and algorithms)
for which the average case analysis is relevant 
\cite{Barg,Spielman}. Let us focus,
for sake of concreteness, on the decoding problem. Codewords are sequences
of symbols with some built-in redundancy. If we consider the case of
linear codes on a binary alphabet, this redundancy can be implemented as a set
of linear constraints. In practice, a codeword is a vector 
$\utx\in \{0,1\}^N$ (with $N\gg 1$) which satisfies the equation

\begin{eqnarray}
{\mathbb H}\, \utx = \ut0\;\;\;\; ({\rm mod}\;\;\; 2)\, ,
\label{ParityCheckMatrix}
\end{eqnarray}

where ${\mathbb H}$ is an $M\times N$ binary matrix 
({\it parity check matrix}). Each one of the $M$ linear equations
involved in Eq. (\ref{ParityCheckMatrix}) is called a {\it parity check}.
This set of equation can be represented graphically by a {\it Tanner graph},
cf. Fig. \ref{Tanner}. This is a bipartite  graph 
highlighting the relations between the variables ${\tt x}_i$ and the constraints
(parity checks) acting on them.  
The decoding problem consists in finding, among the solutions of 
Eq. (\ref{ParityCheckMatrix}), the ``closest'' one $\utx_{\rm d}$ to 
the output $\utx_{\rm out}$ of
some communication channel. This problem is, in general, NP-hard 
\cite{Berlekamp}.

The precise meaning of ``closest'' depends upon the nature of the 
communication channel. Let us make two examples:
\begin{itemize}
\item The binary symmetric channel (BSC). In this case the output of 
the communication channel $\utx_{\rm out}$ is a codeword, i.e. a solution of 
(\ref{ParityCheckMatrix}), in which a fraction $p$ of the entries has 
been flipped. ``Closest'' has to be understood in the Hamming-distance
sense. $\utx_{\rm d}$ is the solution of Eq. (\ref{ParityCheckMatrix})
which minimizes the Hamming distance from  $\utx_{\rm out}$.
\item The binary erasure channel (BEC). The output $\utx_{\rm out}$ 
is a codeword
in which a fraction $p$ of the entries has been erased. One has to
find a solution $\utx_{\rm d}$
of Eq. (\ref{ParityCheckMatrix}) which is compatible with 
the remaining entries. Such a problem has a {\it unique}
solution for small enough erasure probability $p$.
\end{itemize}
There are two sources of randomness in the decoding problem: $(i)$ the matrix
${\mathbb H}$ which defines the code is usually drawn from
some random {\it ensemble}; $(ii)$ the received message which is distributed
according to some probabilistic model of the communication channel
(in the two examples above, the bits to be flipped/erased were 
chosen randomly). Unlike many other combinatorial problems, 
there is therefore a ``natural'' probability distribution defined
on the instances. Average case analysis with respect to this
distribution is of great practical relevance.

Recently, amazingly good performances have been obtained by using
low-density parity check (LDPC) codes \cite{Chung}. LDPC codes 
are defined by parity check matrices ${\mathbb H}$ which are large and sparse.
As an example we can consider Gallager {\it regular} codes 
\cite{GallagerThesis}. In this case ${\mathbb H}$
is chosen with flat probability distribution within
the family of matrices having  $l$ ones per column,
and $k$ ones per row.
These are decoded using a suboptimal linear-time algorithm known as 
``belief-propagation'' or ``sum-product'' algorithm 
\cite{GallagerThesis,Pearl}.
This is an iterative algorithm which takes advantage of the locally tree-like
structure of the Tanner graph, see Fig. \ref{Tanner}, for LDPC codes.
After $n$ iterations it incorporates the information 
conveyed by the variables up to distance $n$ from the one to be decoded.
This can be done in a recursive fashion allowing for linear-time
decoding.

Belief-propagation decoding shows a striking threshold phenomenon
as the noise level $p$ crosses some critical (code-dependent) value $p_d$.
While for $p<p_d$ the transmitted codeword is recovered with 
high probability, for $p>p_d$ decoding will fail almost always. 
The threshold noise $p_d$ is, in general, smaller
than the threshold $p_c$ for optimal decoding (with unbounded 
computational resources).

The rigorous analysis of Ref.~\cite{RichardsonUrbankeIntroduction}
allows a precise determination of the critical noise $p_d$ under
quite general circumstances. Nevertheless some important theoretical
questions remain open: 
Can we find some smarter linear-time algorithm whose threshold
is greater than $p_d$? Is there any ``intrinsic'' (i.e. algorithm independent)
characterization of the threshold phenomenon
taking place at $p_d$? 
As a first step towards the answer to these questions, 
Ref.~\cite{DynamicCodes} explored the dynamics of local optimization
algorithms by using statistical mechanics techniques.
The interesting point is that ``belief propagation'' is by no means a local
search algorithm.

For sake of concreteness, we shall focus on the binary erasure channel.
In this case we can treat decoding as a combinatorial
optimization problem within the space of bit sequences of length $Np$
(the number of erased bits, the others being fixed by the received
message).
The function to be minimized is the {\it energy density}

\begin{eqnarray}
\epsilon(\utx) = \frac{2}{N}d_H({\mathbb H}\utx,\ut0)\, ,\label{CostFunction}
\end{eqnarray}

where we denote as $d_{H}(\utx_1,\utx_2)$ the Hamming distance between
two vectors $\utx_1$ and $\utx_2$, and we 
introduced the normalizing factor for future convenience.
Notice that both arguments of $d_H(\cdot,\cdot)$ in Eq. (\ref{CostFunction})
are vectors in $\{0,1\}^{M}$.

We can define the $R$-neighborhood of a given sequence $\utx$ as the set
of sequences $\utz$ such that $d_{\rm H}(\utx,\utz)\le R$, 
and we call $R$-stable states the bit sequences which are optima
of the decoding problem within their $R$-neighborhood.

\begin{figure}
\centerline{
\includegraphics[height=220pt,angle=-90] {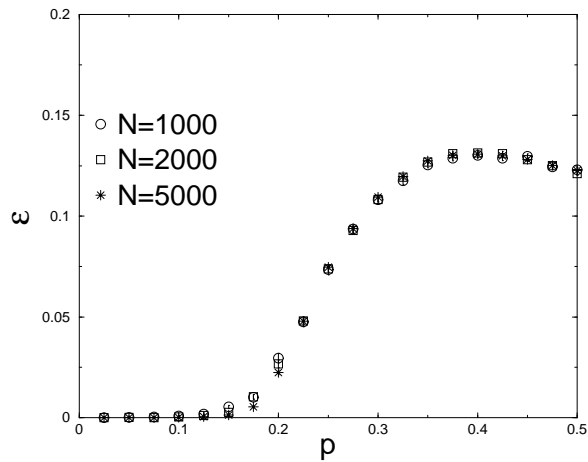}
}
\caption{The $(6,3)$ Gallager code decoded by local search with 
$1$-neighborhoods. At each time step, the algorithm looks for a
 bit (among the ones uncorrectly received) 
such that flipping it decreases the cost function (\ref{CostFunction}). 
We plot the average number of violated parity checks (multiplied by $2/N$) 
after the algorithm halts, as a
function of the erasure probability $p$.}
\label{Glauber}
\end{figure}

One can easily invent local search algorithms
\cite{papadimi} for the decoding problem
which use the $R$-neighborhoods. The algorithm start from a random
sequence and, at each step, optimize it within its $R$-neighborhood.
This algorithm is clearly suboptimal and halts on $R$-stable states.
Let us consider, for instance, a $(k=6,l=3)$ regular code  
and decode it  by local search in $1$-neighborhoods. 
In Fig.~\ref{Glauber} we report the resulting energy density $\epsilon$ after
the local search algorithm halts, as a function of the erasure
probability $p$. We averaged over 100 different realizations of the
noise and of the matrix ${\mathbb H}$. For sake of comparison
we recall that the threshold for belief-propagation decoding
is $p_d\approx 0.429440$ \cite{RichardsonUrbankeIntroduction}, 
while the threshold for optimal 
decoding is at $p_c\approx 0.488151$ \cite{DynamicCodes}.
It is evident that local search by $1$-neighborhoods performs quite poorly.

A natural question is whether (and how much), these performances
are improved by increasing $R$. 
It is therefore quite natural to study {\it metastable} states. 
These are $R$-stable states
for any $R = o(N)$\footnote{We use the standard notation
$f_N=o(N)$ if $\lim_{N\to\infty}f_N/N = 0$.}. There exists no 
completely satisfying definition of such states: here we
shall just suggest a possibility among others. 
The tricky point is that we do not know how to compare
$R$-stable states for different values of $N$. 
This forbids us to make use of the above asymptotic statement.
One possibility is to count without really defining them.
This can be done, at least in principle, by counting $R$-stable
states, take the $N\to\infty$ limit and, at the end, the $R\to\infty$ 
limit\cite{giu}.
On physical grounds, we expect $R$-stable states to be exponentially
numerous. In particular, if we call ${\cal N}_{R}(\epsilon)$ the number of
$R$-stable states taking a value $\epsilon$ of the cost function 
(\ref{CostFunction}), we have 

\begin{eqnarray}
{\cal N}_{R}(\epsilon)\sim \exp\{N S_R(\epsilon)\}\, .
\end{eqnarray}

We can therefore define the so called 
(physical) complexity $\Sigma(\epsilon)$ as follows,

\begin{eqnarray}
\Sigma(\epsilon) \equiv \lim_{R\to\infty} S_R(\epsilon)\, .
\end{eqnarray}

Roughly speaking we can say that the number of metastable states is
$\exp\{N\Sigma(\epsilon)\}$.
Of course there are several alternative ways of taking the 
limits $R\to\infty$, $N\to\infty$, and we do not yet have a proof 
that these procedures give the same result for $\Sigma(\epsilon)$.
Nevertheless it is quite clear that the existence of an exponential number of
metastable states should affect dramatically the behavior of local search
algorithms.

\begin{figure}
\centerline{
\includegraphics[height=220pt,angle=-90] {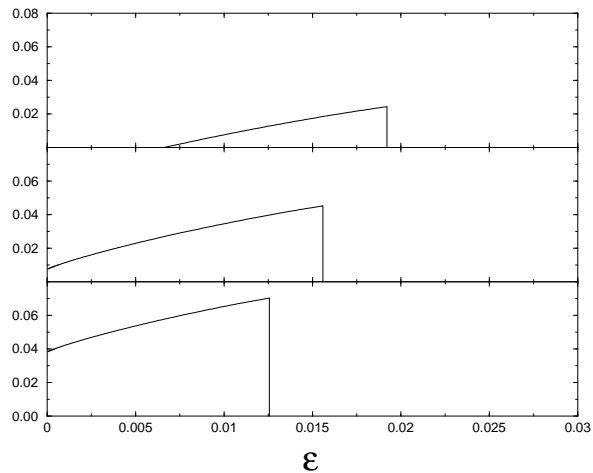}}
\caption{The complexity $\Sigma(\epsilon)$ of a $(6,3)$ code on the BEC,
for (from top to bottom) 
$p=0.45$ (below $p_c$), $p = 0.5$, and $p=0.55$ (above $p_c$).
Recall that $\Sigma(\epsilon)$ is positive only above $p_d\approx 0.429440$.}
\label{Complexity}
\end{figure}

Statistical mechanics methods \cite{DynamicCodes} 
allows to determine the complexity $\Sigma(\epsilon)$ \cite{Remi}. 
In ``difficult'' cases (such as for error-correcting codes), 
the actual computation may involve some approximation,
e.g.  the use of a variational Ansatz. Nevertheless the outcome
is usually quite accurate. 
In Fig. \ref{Complexity} we consider a $(6,3)$ regular code on the
binary erasure channel.
We report the resulting complexity 
for three different values of the erasure probability $p$.
The general picture is as follows. Below $p_d$ there is no metastable
state, except the one corresponding to the correct
codeword. Between $p_d$ and $p_c$ there is an exponential number
of metastable states with energy density belonging to the interval
$\epsilon_{GS}<\epsilon<\epsilon_D$ ($\Sigma(\epsilon)$ is strictly positive
in this interval).
Above $p_c$, $\epsilon_{GS}=0$.
The maximum of $\Sigma(\epsilon)$ is always at $\epsilon_D$.

\begin{figure}
\centerline{
\includegraphics[height=220pt,angle=-90] {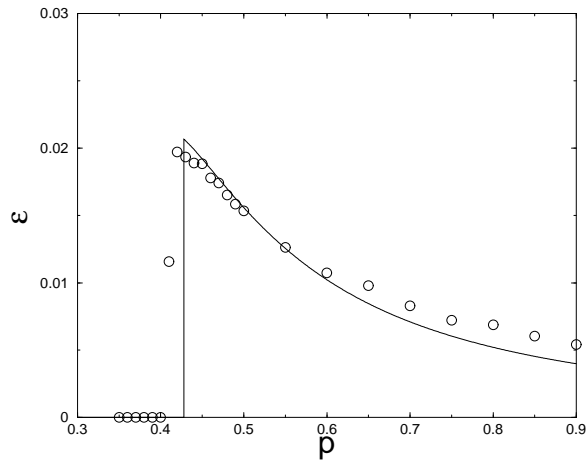}}
\caption{The $(6,3)$ LDPC code on the BSC decoded by simulated annealing.
The circles give the number of violated checks in the resulting sequence.
The continuous line is the analytical result for the typical energy
density of metastable states ($\epsilon_D$ in Fig. \ref{Complexity}).}
\label{Annealing}
\end{figure}

The above picture tell us that any local algorithm will run into
difficulties above $p_d$. In order to confirm this picture, 
the authors of Ref.~\cite{DynamicCodes} made
some numerical computations using simulated annealing as  decoding
algorithm for
 quite large codes ($N=10^4$ bits).  For each value of
$p$, we start the simulation fixing a fraction $(1-p)$ of spins to
$\sigma_i = +1$ (this part will be kept fixed all along the run).  The
remaining $p N$ spins are the dynamical variables we change during the
annealing in order to try to satisfy all the parity checks.  The
energy of the system counts the number of unsatisfied parity checks.

The cooling schedule has been chosen in the following way: $\tau$
Monte Carlo sweeps (MCS)~\footnote{Each Monte Carlo sweep consists in
$N$ proposed spin flips. Each proposed spin flip is accepted or not
accordingly to a standard Metropolis test.}
at each of the 1000 equidistant temperatures
between $T=1$ and $T=0$.  The highest temperature is such that the
system very rapidly equilibrates.  
Typical values for $\tau$ are from 1 to $10^3$.

Notice that, for any fixed cooling schedule, the computational complexity of
the simulated annealing method is linear in $N$.  Then we expect it to be
affected by metastable states of energy $\epsilon_D$, which are
present for $p>p_d$: the energy relaxation should be strongly reduced
around $\epsilon_D$ and eventually be completely blocked.
 Some results are plotted in
Fig. \ref{Annealing} together with the theoretical prediction for
$\epsilon_D$. The good agreement confirm our picture: the algorithm 
gets stucked in metastable states, which have, in the great majority
of cases, energy density $\epsilon_D$.

Both ``belief propagation'' and local search algorithms fail to
decode correctly between $p_d$ and $p_c$. This leads naturally to the
following conjecture: no linear time algorithm
can decode in this regime of noise. The (typical case) 
computational complexity changes from
being linear below $p_d$ to superlinear above $p_d$. In the case of
the binary erasure channel it remains polynomial between $p_d$ and
$p_c$ (since optimal decoding can be realized with linear algebra methods). 
However it is plausible that for a general channel it becomes 
non-polynomial.

\subsection{Gradient descent and XORSAT}
\label{gradxor}

In this section the local procedure we consider is gradient descent (GD).
GD is defined as follows. {\bf (1)} Start from an initial randomly chosen
configuration of the variables. Call $E$ the number of unsatisfied
clauses.  {\bf (2)} If $E=0$ then stop (a solution is found). Otherwise,
pick randomly one variable, say $x_i$, and compute the number $E'$ of
unsatisfied clauses when this variable is negated; if $E'\ge E$
then accept this change {\em i.e.} replace $x_i$ with $\bar x_i$ and
$E$ with $E'$; if $E' >E$, do not do anything. Then go to step 2.
The study of the performances of GD to find the minima of cost
functions related to statistical physics models has recently
motivated various studies\cite{gdphys,gdproof}. Numerics indicate that
GD is typically able to solve random 3-SAT instances with ratios $\alpha
< 3.9$ \cite{suedois,cavi} close to the onset of clustering
\cite{varia,cavi,par}. We shall rigorously show below that this is not so
for 3-XORSAT.

\begin{figure}
\centerline{\includegraphics[scale=0.5,angle=0]{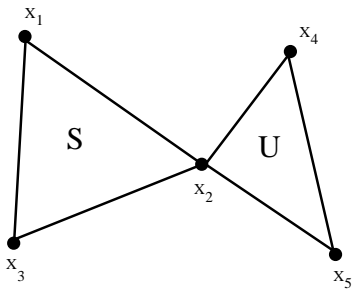}}
\caption{Graphical representation of the XORSAT instance with two
clauses involving variables
$x_1,x_2,x_3$ and $x_2,x_4,x_5$. Each clause or equation
is represented by a plaquette whose vertices are the attached variables.
When the variables are assigned some values, the clauses can
be satisfied (S) or unsatisfied (U).  }
\label{xorgr}
\end{figure}

Let us apply GD to an instance of XORSAT. The instance has a graph
representation explained in Fig.~\ref{xorgr}. Vertices are in 
one--to--one correspondence with variables. A clause is 
fully represented 
by a plaquette joining three variables and a Boolean label
equal to the number of negated variables it contains modulo 2
(not represented on Fig.~\ref{xorgr}). 
Once a configuration of the variables is chosen, each plaquette
may be labelled by its status, S or U,  whether the associated
clause is respectively satisfied or unsatisfied. A fundamental
property of XORSAT is that each time a variable is changed, {\em i.e.}
its value is negated, the clauses it belongs to change status too. 

\begin{figure}
\centerline{\includegraphics[scale=0.5,angle=0]{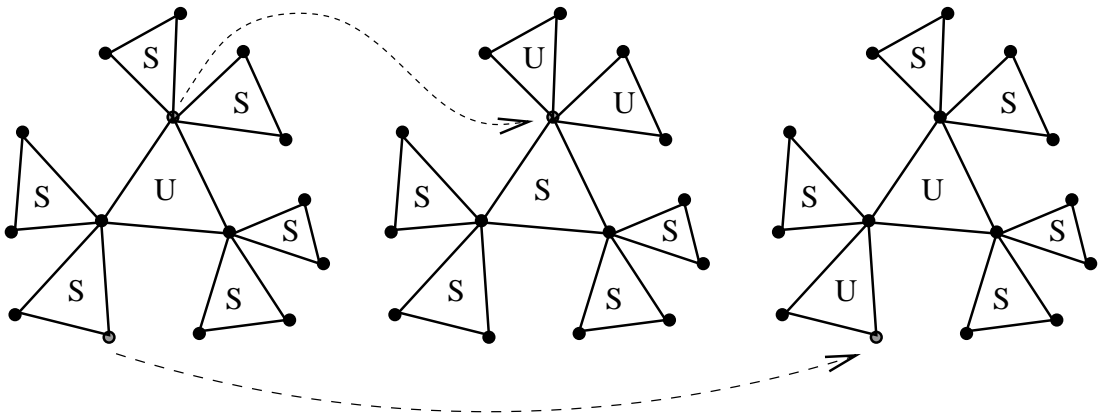}}
\caption{A blocked island (left) is an instance of 7 clauses (1 central, 6
peripherical) with 
variables such that the central plaquette is unsatisfied and all
peripherical plaquettes are satisfied. Inversion of any variable (grey vertex)
increases the number of unsatisfied clauses by 1, be it attached to the central
(middle) or to a peripherical (right) plaquette.}
\label{bi}
\end{figure}

This property makes the analysis of some properties of GD easy.
Consider the hypergraph made of 15 vertices and 7 plaquettes in 
Fig.~\ref{bi},
and suppose the central plaquette is violated (U) while all other 
plaquettes are satisfied (S). The number of unsatisfied clauses
is $E=1$. Now run GD on this special instance of XORSAT.
Two cases arise, symbolized in Fig.~\ref{bi}, whether the vertex
attached to the variable to be flipped 
belongs, or not, to the central plaquette.
It is an easy check that, in both cases, $E'=2$ and the change is not
permitted by GD. The hypergraph of Fig.~\ref{bi} will be called hereafter
island. When the status of the plaquettes is U for the central one
and S for the other ones, the island is called blocked. Though the 
instance of the XORSAT problem encoded by a blocked island is obviously
satisfiable (think of negating at the same time one variable attached
to a vertex $V$ of the central plaquette and one variable in each of 
the two peripherical plaquettes joining the central plaquette at $V$),
GD will never be able to find a solution and will be blocked forever
in the local minimum with height $E=1$.

The purpose of this section is to show that this situation typically
happens for random instances of XORSAT. More precisely, while almost all
instances of XORSAT with a ratio of clauses per variables smaller than
$\alpha \simeq 0.918$ have a lot of solutions, GD is almost never able to find 
one. Even worse, the number of violated clauses reached by GD
is bounded from below by $\Psi (\alpha) \, N$ where
\begin{equation} \label{teh}
\Psi (\alpha) = \frac{729}{1024} \, \alpha^7 \, e^{-45\, \alpha} \qquad .
\end{equation}
In other words, the number of clauses remaining unsatisfied at the
end of a typical GD run is of the order of $N$.
Our demonstration, inspired from \cite{gdproof}, 
is based on the fact that, with high probability, a
random instance of XORSAT contains a large number of blocked islands
of the type of Fig.~\ref{bi}.

To make the proof easier, we shall study the following fixed clause 
probability ensemble. Instead of imposing
the number of clauses to be equal to $M (=\alpha N)$, any triplet $\tau$ 
of three vertices (among $N$) is allowed to carry a plaquette  with
probability $\mu = \alpha N/{N \choose 3} = 6\alpha /N^2 + O(1/N^3)$. Notice
that this probability ensures that, on the average, the number of
plaquettes equals $\alpha N$. 
Let us now draw a hypergraph with this distribution.
For each triplet $\tau$ of vertices, we define $I_{\tau}=1$ if
$\tau$ is the center of a  island, 0 otherwise. We shall show 
that the total number of  islands, 
$I = \sum _{\tau} I_{\tau}$, is highly concentrated
in the large $N$ limit, and calculate its average value.

The expectation value of $I_{\tau}$ is equal to
\begin{equation} \label{expitau}
E [I_{\tau} ] = \frac{(N-3)\times (N-4) \times \ldots \times 
(N-13)\times (N-14)}{8\times 8\times 8} \times \mu ^A \, (1-\mu) ^B 
\ ,
\end{equation}
where $A=7$ is the number of plaquettes in the island, and
\begin{equation}
B = {N \choose 3} - {N -15 \choose 3} - 7 \qquad ,
\end{equation}
is the number of triplets with at least one vertex among the set of 15 vertices
of the island that do not carry plaquette. The significance of the
terms in Eq.~(\ref{expitau}) is transparent. The central
triplet $\tau$ occupying three vertices, we choose 2 vertices among $N-3$
to draw the first peripherical plaquette of the island, then other 2
vertices among $N-5$ for the other peripherical plaquette having a common
vertex with the latter. The order in which these two plaquettes are
built does not matter and a factor $1/2$ permits us to avoid double
counting. The other four peripherical plaquettes have multiplicities 
calculable in the same way (with less and less available vertices). 
The terms in 
$\mu$ and $1-\mu$ correspond to the probability that such a 7 plaquettes
configuration is drawn on the 15 vertices of the island, and is disconnected
from the remaining $N-15$  vertices. The expectation value of the
number $i=I/N$ of  islands per vertex thus reads,
\begin{equation} \label{eq89}
\lim _{N\to\infty} E[ i] =  \lim _{N\to\infty} \frac
1N \, {N \choose 3}\, E[I_{\tau}] = \frac{729}{8} \, \alpha^7 \, e^{-45\, 
\alpha} \quad .
\end{equation}
Chebyshev's inequality can be used to show that $i$ is concentrated 
around its above average value. Let us calculate the second moment
of the number of  islands, $E[I^2]= \sum _{\tau , \sigma }
E[ I_{\tau} I_{\sigma}] $. Clearly,  $E[ I_{\tau} I_{\sigma}]$ depends
only on the number $\ell=0,1,2,3$ of vertices common to triplets $\tau$
and $\sigma$. It is obvious that no two triplets of vertices
can be centers of  islands when they have $\ell=1$ or $\ell=2$ common
vertices. If $\ell=3$, $\tau=\sigma$ and $E_{\ell =0} \equiv 
E[ I_{\tau} ^2]=E[ I_{\tau}]$ has been calculated above. 
For $\ell=0$, a similar calculation gives
\begin{eqnarray}
E_{\ell =0} &=& \frac{(N-6) (N-7) ... (N-29)}{2^{18}} \, \mu ^{14}
(1-\mu) ^{B'} \\ \nonumber
B' &=& {N \choose 3} - {N -30 \choose 3} - 14 
\qquad .
\end{eqnarray}
Finally, we obtain
\begin{equation}
E[i^2] = \frac 1{N^2} \left[ {N \choose 3} E_{\ell =3} +
{N \choose 3}{N -3\choose 3} E_{\ell =0} \right] =
E[i]^2 + O\left( \frac 1N \right) 
\quad .
\end{equation} 
Therefore the variance of $i$ vanishes and $i$ is, with high
probability, equal to its average value given by (\ref{eq89}).
To conclude, an island has a probability $1/2^7=1/128$ to be blocked
by definition. Therefore the number (per vertex) 
of blocked islands in a random
XORSAT instance with ratio $\alpha$ is almost surely equal to $\Psi (
\alpha)$
given by Eq.~(\ref{teh}). Since each blocked island has one unsatisfied
clause, this is also a lower bound to the number of violated
clauses per variable. Notice however that $\Psi (\alpha )$ is very small
and bounded from above by $1.5\ 10^{-9}$ over the range of interest, 
$0<\alpha<0.918$. Therefore, one would in principle need to deal
with billions of variables not to reach solutions and be in the true
asymptotic regime of GD. 

The proof is easily generalizable to gradient descent with more than
one look ahead. To extend the notion of blocked island to
the case where GD is allowed to invert $R$, and not only 1, variables
at a time, it is sufficient to have $R+1$, and not 2, peripherical
plaquettes attached to each vertex of the central plaquette. The
calculation of the lower bound $\Psi (\alpha ,R)$ to the number of violated 
clauses (divided by $N$) reached by GD is straightforward and not reproduced
here. As a consequence, GD, even with $R$ simultaneous
flips permitting to overcome local barriers, remains almost surely
trapped at an extensive (in $N$) level of violated clauses for any
finite $R$. Actually the lower bound $\Psi (\alpha ,R) N$ tends
to zero only if $R$ is of the order of $\log N$.

We stress that the statistical physics calculation of
physical `complexity' $\Sigma$ (see Sec.~\ref{CodeSection}) predicts there
is no metastable states for $\alpha < 0.818$\cite{xorsat1},
while GD is almost surely 
trapped by the presence of blocked islands for any $\alpha >0$. 
This apparent discrepancy comes
from the fact that GD is sensible to the presence of configurations
blocked for finite $R$, while the physical `complexity' considers states
metastable in the limit $R\to\infty$ only\cite{giu}.

\subsection{The WalkSAT procedure}
\label{WalkSatSection}

The Pure Random WalkSAT (PRWSAT) algorithm for solving $K$-SAT is
defined by the following rules\cite{papa2}.

\begin{enumerate}
\item Choose randomly a configuration of the Boolean variables.
\item If all clauses are satisfied, output ``Satisfiable''.
\item If not, choose randomly one of the unsatisfied clauses, and one
among the $K$ variables of this clause.  Flip (invert) the chosen
variable.  Notice that the selected clause is now satisfied, but the
flip operation may have violated other clauses which were previously
satisfied.
\item Go to step 2, until a limit on the number of flips fixed
beforehand has been reached. Then Output ``Don't know''.
\end{enumerate}
 
What is the output of the algorithm? Either ``Satisfiable'' and a
solution is exhibited, or ``Don't know'' and no certainty on the
status of the formula is achieved. Papadimitriou introduced this
procedure for $K=2$, and showed that it solves with high probability
any satisfiable 2-SAT instance in a number of steps (flips) of the
order of $N^2$\cite{papa2}. Recently Sch\"oning was able to prove the
following very interesting result for 3-SAT\cite{scho}. Call `trial' a
run of PRWSAT consisting of the random choice of an initial
configuration followed by $3\times N$ steps of the procedure.  If none
of $T$ successive trials on a given instance has been successful (has
provided a solution), then the probability that this instance is
satisfiable is lower than $\exp( - T \times (3/4)^N)$. In other words,
after $T\gg (4/3)^N$ trials of PRWSAT, most of the configuration space
has been `probed': if there were a solution, it would have been found.
Though this local search algorithm is not complete, the uncertainty on
its output can be made as small as desired and it can be used to prove
unsatisfiability (in a probabilistic sense).

Sch\"oning's bound is true for any instance. Restriction to special
input distributions allows to strengthen this result. Alekhnovich and
Ben-Sasson showed that instances drawn from the random
3-Satisfiability ensemble described above are solved in polynomial
time with high probability when $\alpha$ is smaller than
$1.63$\cite{ben}.

\subsubsection{Behaviour of the algorithm}
\begin{center}
\begin{figure}
A\epsfig{file=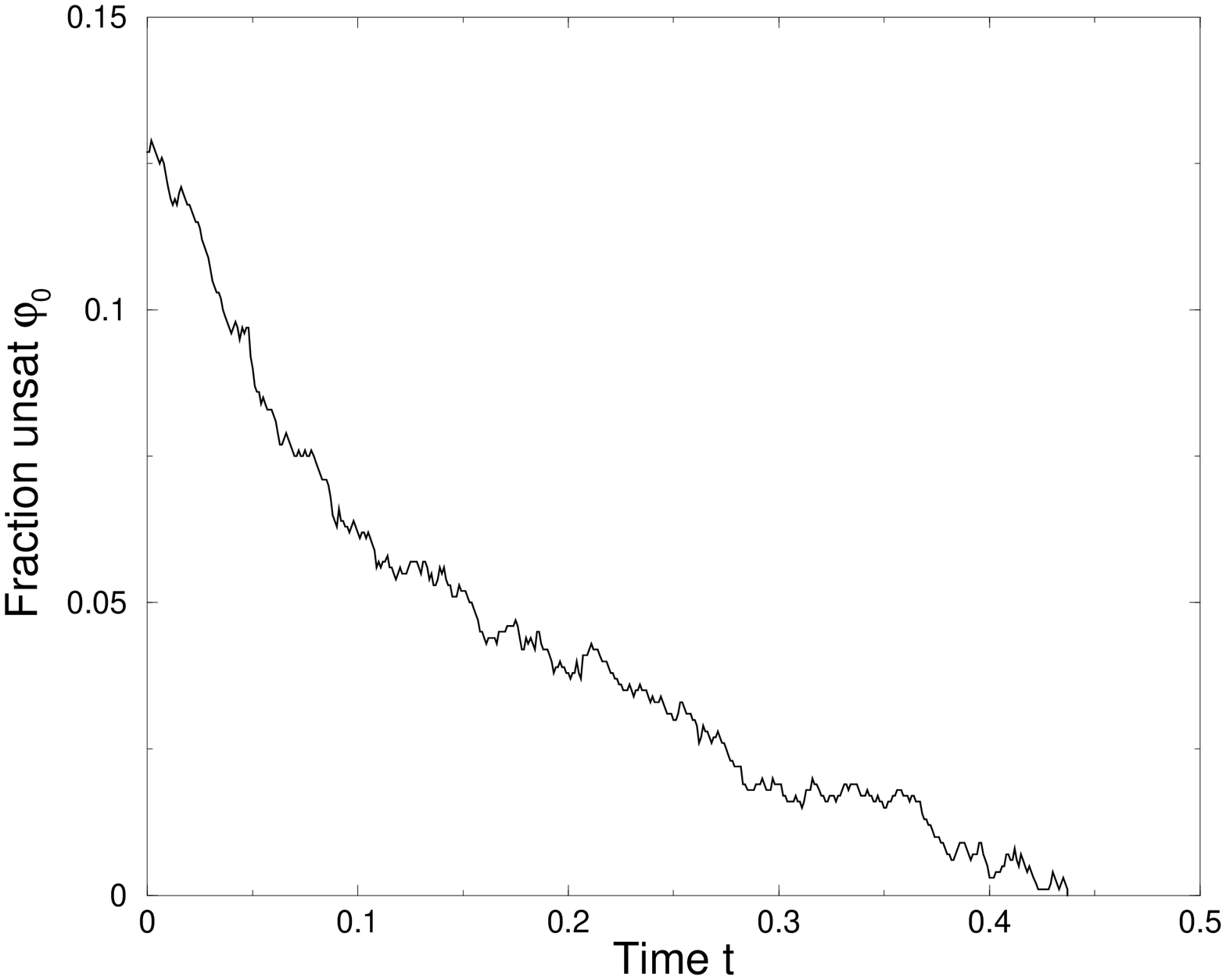,angle=-90,width=7cm} \hskip 1cm
B\epsfig{file=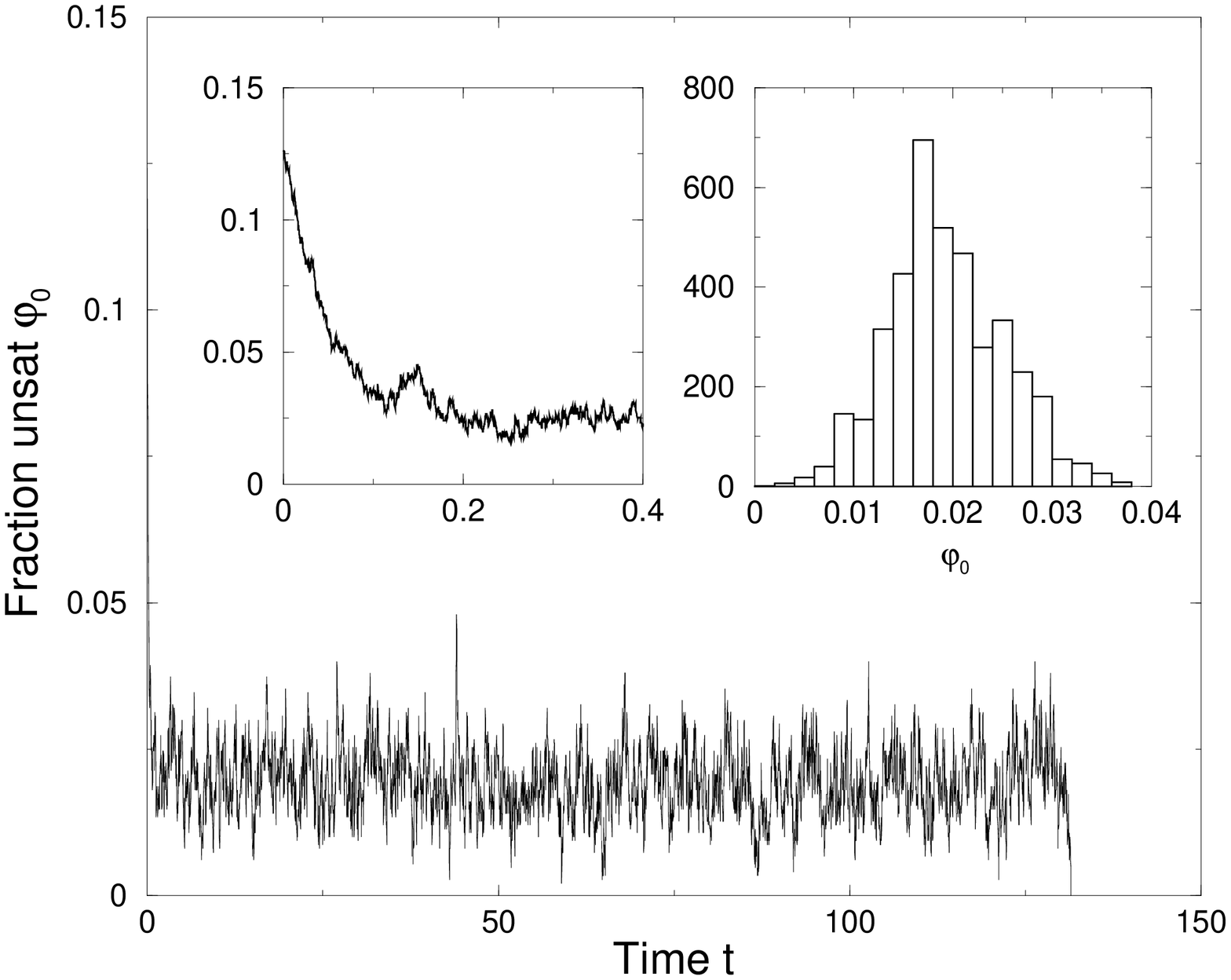,angle=-90,width=7cm} \vskip .5cm
\caption{Fraction $\varphi _0$ of unsatisfied clauses as a function of time
$t$ (number of flips over $M$) during the action of PRWSAT on two randomly drawn instances of
3-SAT with ratios $\alpha =2$ ({\bf A}) and $\alpha =3$ ({\bf B}) with
$N=500$ variables.  Note the difference of time scales between the two
figures. Insets of figure B: left: blow up of the initial relaxation
of $\varphi_0$, taking place on the $O(1)$ time scale as in ({\bf A});
right: histogram $p_{500} (\varphi _0 )$ of the fluctuations of
$\varphi _0$ on the plateau $1\le t\le 130$. }
\label{wsat_phen}
\end{figure}
\end{center}

In this section, we briefly sketch the behaviour of PRWSAT, as seen
from numerical experiments~\cite{Parkes} and the analysis of
\cite{notrewsat,leurwsat}. A dynamical threshold $\alpha _d $ ($\simeq
2.7$ for 3-SAT) is found, which separates two regimes:
\begin{itemize}
\item for $\alpha < \alpha _d$, the algorithm finds a solution very
quickly, namely with a number of flips growing linearly with the
number of variables $N$.  Figure~\ref{wsat_phen}A shows the plot of
the fraction $\varphi _0$ of unsatisfied clauses as a function of time
$t$ (number of flips divided by $M$) for one instance with ratio
$\alpha=2$ and $N=500$ variables. The curve shows a fast decrease from
the initial value ($\varphi _0(t=0)=1/8$ in the large $N$ limit
independently of $\alpha$) down to zero on a time scale
$t_{res}=O(1)$. Fluctuations are smaller and smaller as $N$ grows.
$t_{res}$ is an increasing function of $\alpha$.  This {\em
relaxation} regime corresponds to the one studied by Alekhnovich and
Ben-Sasson, and $\alpha _d > 1.63$ as expected\cite{ben}.

\item for instances in the $\alpha_d < \alpha < \alpha _c$ range, the
initial relaxation phase taking place on $t=O(1)$ time scale is not
sufficient to reach a solution (Fig.~\ref{wsat_phen}B). The fraction
$\varphi_0$ of unsat clauses then fluctuates around some plateau value
for a very long time.  On the plateau, the system is trapped in a {\em
metastable} state.  The life time of this metastable state (trapping
time) is so huge that it is possible to define a (quasi) equilibrium
probability distribution $p_N(\varphi _0)$ for the fraction
$\varphi_0$ of unsat clauses.  (Inset of Fig.~\ref{wsat_phen}B). The
distribution of fractions is well peaked around some average value
(height of the plateau), with left and right tails decreasing
exponentially fast with $N$, $p_N(\varphi _0) \sim \exp ( N \bar \zeta
(\varphi_0))$ with $\bar \zeta \le 0$.  Eventually a large negative
fluctuation will bring the system to a solution ($\varphi
_0=0$). Assuming that these fluctuations are independant random events
occuring with probability $p_N(0)$ on an interval of time of order
$1$, the resolution time is a stochastic variable with exponential
distribution. Its average is, to leading exponential order, the
inverse of the probability of resolution on the $O(1)$ time scale:
$[t_{res}] \sim \exp (N \zeta)$ with $\zeta = - \bar \zeta (0)$.
Escape from the metastable state therefore takes place on
exponentially large--in--$N$ time scales, as confirmed by numerical
simulations for different sizes.  Sch\"oning's result\cite{scho} can
be interpreted as a lower bound to the probability $\bar \zeta (0)>
\ln (3/4)$, true for any instance.
\end{itemize}

\begin{figure}
\begin{center}
\epsfig{file=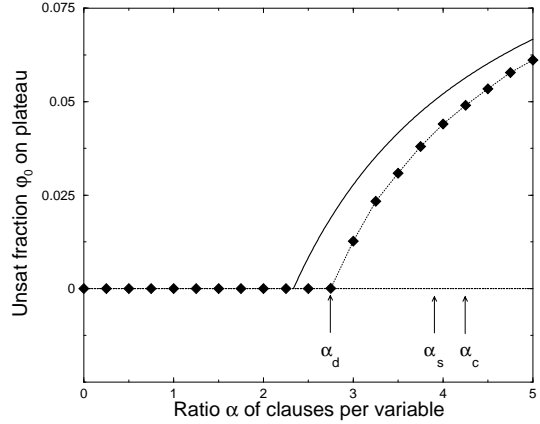,angle=-90,width=7cm} \vskip .5cm
\end{center}\caption{Fraction $\varphi _0$ of unsatisfied clauses on 
the metastable plateau of PRWSAT on 3-SAT as a function of the ratio $\alpha$ of clauses
per variable. Diamonds are the output of numerical experiments, and
have been obtained through average of data from simulations at a given
size $N$ (nb. of variables) over 1,000 samples of 3-SAT, and
extrapolation to infinite sizes (dotted line serves as a guide to the
eye).  The ratio at which $\varphi _0$ begins being positive, $\alpha
_d \simeq 2.7$, is smaller than the thresholds $\alpha _s\simeq 3.9$
and $\alpha _c\simeq 4.3$ above which solutions gather into distinct
clusters and instances have almost surely no solution respectively.
The full line is the prediction of the Markovian approximation of
Section~\ref{wsat_expphase}.}
\label{wsat_plateau}
\end{figure}

The plateau energy, that is, the fraction of unsatisfied clauses
reached by PRWSAT on the linear time scale is plotted on
Fig.~\ref{wsat_plateau}.  Notice that the ``dynamic'' critical value
$\alpha_d$ above which the plateau energy is positive (PRWSAT stops
finding a solution in linear time) is strictly smaller than the
``static'' ratio $\alpha_c$, where formulas go from satisfiable with
high probability to unsatisfiable with high probability.  In the
intermediate range $\alpha_d < \alpha <\alpha_c$, instances are almost
surely satisfiable but PRWSAT needs an exponentially large time to
prove so.  Interestingly, $\alpha _d$ and $\alpha_c$ coincides for
2-SAT in agreement with Papadimitriou's
result\cite{papa2}. Furthermore, the dynamical transition is
apparently not related to the onset of clustering taking place at
$\alpha _s \simeq 3.9$.

\subsubsection{Results for the linear phase $\alpha < \alpha _d$}

\begin{figure}
\begin{center}
\epsfig{file=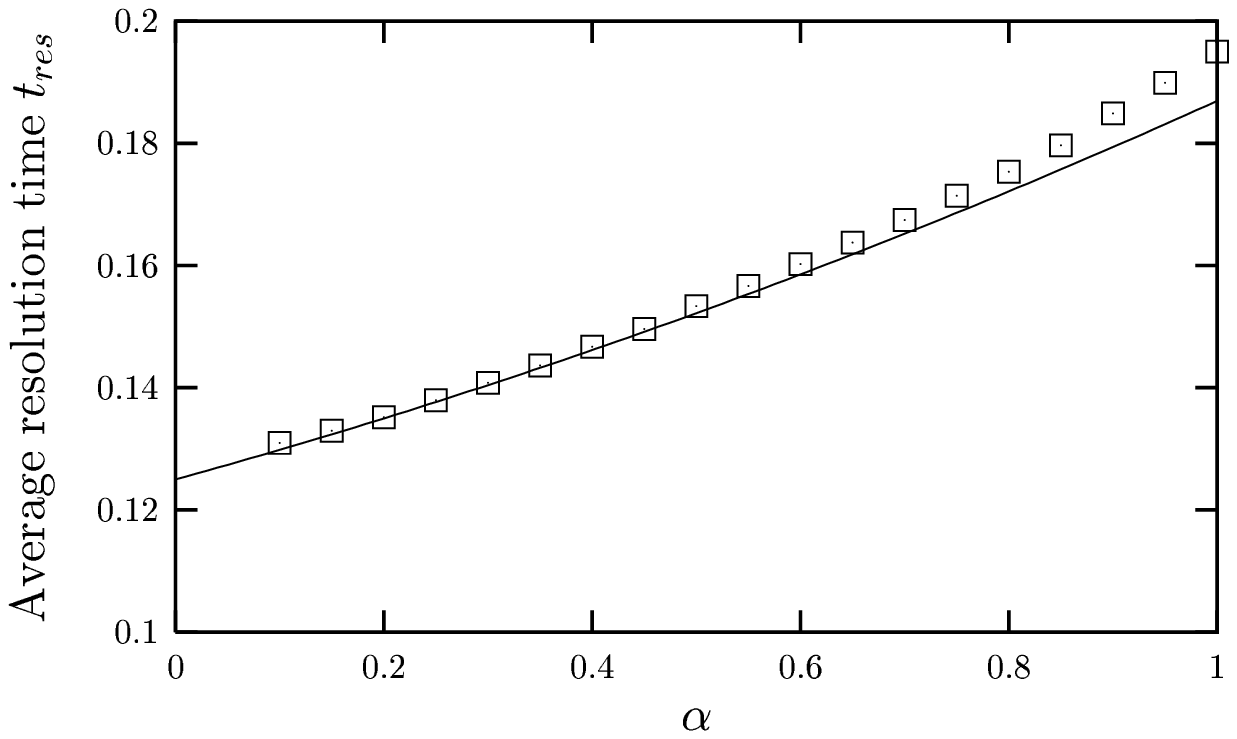,width=7cm}
\end{center}
\caption{Average resolution time $t_{res}(\alpha , 3)$ for PRWSAT on 3-SAT. Symbols: numerical simulations, averaged over $1,000$ runs for
$N=10,000$.  Solid line: prediction from the cluster expansion
(\ref{dev_cluster_tresK}).}
\label{wsat_fig_tres_q1}
\end{figure}

When PRWSAT finds easily a solution, the number of steps it requires
is of the order of $N$, or equivalently, $M$.  Let us call
$t_{res}(\alpha,K)$ the average of this number divided by the number
of clauses $M$.  By definition of the dynamic threshold, $t_{res}$
diverges when $\alpha \to \alpha_d^-$. 
Assuming that $t_{res}(\alpha ,K)$ can be expressed as a series of 
powers of $\alpha$, we find the following expansion\cite{notrewsat}
\begin{equation}
t_{res} (\alpha , K)= \frac{1}{2^K} + \frac{K(K+1)}{K-1}
\frac{1}{2^{2K+1}} \, \alpha + \frac{4K^6 + K^5 +6 K^3 -10 K^2 + 2
K}{3(K-1)(2K-1)(K^2-2)} \frac{1}{2^{3K+1}} \, \alpha^2 + O(\alpha^3)
\quad .\label{dev_cluster_tresK}
\end{equation}
around $\alpha=0$. As only a finite number of terms in this
expansion have been computed, 
we do not control its radius of convergence, yet as shown
in Fig.~\ref{wsat_fig_tres_q1} the numerical experiments provide
convincing evidence in favour of its validity. 

The above calculation is based on two facts. First, for
$\alpha < 1/(K(K-1))$ the instance under consideration splits into
independent subinstances (involving no common variable) 
that contains a number of variables of the order 
of $\log N$ at most. Moreover,
the number of the connected components containing $m$ clauses,
computed with probabilistic arguments very similar to those of
Section~\ref{gradxor}, contribute to a power expansion in $\alpha$
only at order $\alpha^m$. Secondly, the number of steps the
algorithm needs to solve the instance is simply equal to the sum
of the numbers of steps needed for each of its independent subinstances.
This additivity remains true when one averages over the
initial configuration and the choices done by the algorithm. 

One is then left with the enumeration of the different subinstances with a
given size and the calculation of the average number of steps for
their resolution. A detailed presentation of
this method has been given in a general case in~\cite{clusters}, and
applied more specifically to this problem in~\cite{notrewsat}; the
reader is referred to these previous works for more details. 
Equation (\ref{dev_cluster_tresK}) is the output of the enumeration of subinstances 
with up to three clauses.

\subsubsection{Results for the exponential phase $\alpha > \alpha _d$}
\label{wsat_expphase}

The above small $\alpha$ expansion does not allow us to investigate the
$\alpha > \alpha _d$ regime.
We turn now to an approximate method more adapted to this situation.

Let us denote by $C$ an assignment of the boolean variables.  
PRWSAT defines a Markov process on the space of the
configurations $C$, a discrete set of cardinality $2^N$. It is a
formidable task to follow the probabilities of all these
configurations as a function of the number of steps $T$ of the
algorithm so one can look for a simpler
description of the state of the system during the evolution of the
algorithm. The simplest, and crucial, quantity to follow is the number
of clauses unsatisfied by the current assignment of the boolean
variables, $M_0(C)$. Indeed, as soon as this value vanishes, the
algorithm has found a solution and stops. 

\begin{figure}
\begin{center}
\epsfig{file=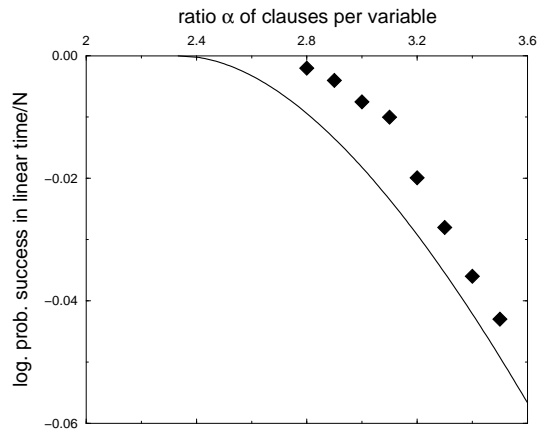,angle=-90,width=7cm}
\end{center}
\caption{Large deviations for the action of PRWSAT on 3-SAT. The logarithm 
$\bar \zeta $ of the probability of successful resolution (over the
linear in $N$ time scale) is plotted as a function of the ratio
$\alpha$ of clauses per variables.  Prediction for $\bar \zeta (\alpha
,3)$ has been obtained within the approximations of
Section~\ref{wsat_expphase}. Diamonds corresponds to (minus) the
logarithm $\zeta$ of the average resolution times (averaged over 2,000
to 10,000 samples depending on the values of $\alpha,N$, divided by
$N$ and extrapolated to $N\to\infty$) obtained from numerical
simulations. Error bars are of the order of the size of the diamond
symbol.  Sch\"oning's bound is $\bar \zeta \ge \ln(3/4) \simeq
-0.288$.}
\label{wsat_fig_zeta}
\end{figure}

A crude approximation consists in assuming that, at each time step
$T$, all configurations with a
given number of unsatisfied clauses are equiprobable, whereas the
Hamming distance between two configurations visited at step $T$ and
$T+k$ of the algorithm is at most $k$. However, the results obtained
are much more sensible that one could fear.
Within this simplification, a Markovian evolution equation for
the probability that $M_0$ clauses are unsatisfied after $T$ steps
can be written. 
Using methods similar to the ones in Section~\ref{DpllSatSection},
we obtain (see \cite{notrewsat} for more
details and \cite{leurwsat} for an alternative way of presenting the
approximation):
\begin{itemize}
\item the average fraction of unsatisfied clauses,
$\varphi_0(t)$, after $T=t\, M$ steps of the algorithm.
For ratios $\alpha > \alpha_d(K)= (2^K - 1) /K$,
$\varphi_0$ remains positive at large times, which means that typically
a large formula will not be solved by PRWSAT, and that the fraction of
unsat clauses on the plateau is 
$\varphi_0(t \to \infty)$. 
The predicted value for $K=3$, $\alpha_d=7/3$, is in
good but not perfect agreement with the estimates from numerical
simulations, around $2.7$. The plateau height, 
$2^{-K}(1-\alpha_d(K)/\alpha)$, is compared to
numerics in Fig.~\ref{wsat_plateau}.

\item the probability $p_N(\varphi _0) \sim \exp ( N \bar \zeta
(\varphi_0))$ that the fraction of unsatisfied clauses is $\varphi _0$.
It has been argued above that
the distribution of resolution times in the $\alpha > \alpha_d$ phase
is expected to be, at leading order, an exponential distribution of
average $e^{N \zeta}$, with $\zeta = -\bar \zeta (0)$. Predictions for
$\bar \zeta (0)$  are plotted and compared to experimental measures
of $\zeta$ in Fig.~\ref{wsat_fig_zeta}. Despite the roughness of our Markovian
approximation, theoretical predictions are in qualitative agreement
with numerical experiments.
\end{itemize}

A similar study of the behaviour of PRWSAT on XORSAT problems has been
also performed in~\cite{notrewsat,leurwsat}, with qualitatively
similar conclusions: there exists a dynamic threshold $\alpha_d$ for
the algorithm, smaller both than the satisfiability and clustering
thresholds (known exactly in this case~\cite{xorsat2}). For low
values of $\alpha$, the resolution time is linear in the size of the
formula; between $\alpha_d$ and $\alpha_c$ resolution occurs on
exponentially large time scales, through fluctuations around a plateau
value for the number of unsatisfied clauses. In the XORSAT case, the
agreement between numerical experiments and this approximate study
(which predicts $\alpha_d=1/K$) is quantitatively better and seems to
improve with growing $K$.

\section{Conclusion and perspectives}

In this article, we have tried to give an overview of the studies that
physicists have devoted to the analysis of algorithms. This
presentation is certainly not exhaustive. Let us mention that use of
statistical physics ideas have permitted to obtain very interesting
results on related issues as number partitioning\cite{mertens},
binary search trees \cite{majum}, learning in neural networks
\cite{nn}, extremal optimization \cite{boe} ...

It may be objected that algorithms are mathematical and well defined 
objects and, as so, should be analysed with rigorous techniques only. 
Though this point of view should ultimately prevail, the current state 
of available probabilistic or combinatorics techniques compared to
the sophisticated nature of algorithms used in computer science make
this goal unrealistic nowadays. We hope the reader is now convinced that
statistical physics ideas, techniques, ... may be of help
to acquire a quantitative intuition or even formulate conjectures on the
average performances of search algorithms.
A wealth of concepts and methods 
familiar to physicists e.g. phase transitions
and diagrams, dynamical renormalization flow, out-of-equilibrium 
growth phenomena, metastability, perturbative approaches... are found 
to be useful to understand
the behaviour of algorithms. It is a simple bet that this list will
get longer in next future and that more and more powerful techniques
and ideas issued from modern theoretical physics will find their place 
in the field.

Open questions are numerous. Variants of DPLL with complex
splitting heuristics, random 
backtrackings\cite{marq} or applied to combinatorial problems
with internal symmetries\cite{liat} would be worth being studied.
As for local search algorithms, it would be very interesting to 
study refined versions of the Pure WalkSAT procedure that alternate
random and greedy steps \cite{defwsat,selmantuning,HoosStutz} to understand
the observed existence and properties of optimal strategies.
One of the main open questions in this context is 
to what extent performances are related to intrinsic
features of the combinatorial problems and not to the details
of the search algorithm\cite{lancas}. This raises the question of 
how the structure of the cost function landscape may induce
some trapping or slowing down of search algorithms\cite{par}.
Last of all, the input distributions of instances we have focused on here
are far from being realistic. Real instances have a lot of structure
which will strongly reflect on the performances of algorithms.
Going towards more realistic distributions or, even better, obtaining
results true for any instance would be of great interest.

\vskip .5cm
{\bf Acknowledgments.}
This work was partly funded by the ACI Jeunes Chercheurs 
``Algorithmes d'optimisation et syst\`emes d\'esordonn\'es quantiques'' 
from the French Ministry of Research.

\end{document}